\documentclass[a4paper,10pt]{report}
\pdfoutput=1
\usepackage[english]{babel}
\usepackage[latin1]{inputenc}
\usepackage{amsfonts,amssymb}
\usepackage{amsmath}
\usepackage{natbib}
\usepackage{graphicx}
\usepackage{color}
\definecolor{gris5}{gray}{0.95}
\numberwithin{equation}{chapter}

\makeatletter
\newcommand{\mythanks}[1]{\hbox{\@textsuperscript{\normalfont#1}}}
\makeatother

\title{\textbf{Dynamic models of residential segregation:\\ Brief review, analytical resolution and study of the introduction of coordination\\
\vspace{1.5cm} Working paper}}
\author{S\'ebastian Grauwin\mythanks{1,2,3,0}, Florence Goffette-Nagot\mythanks{1,4}, Pablo Jensen\mythanks{1,2,3,5}}
\date{\today}

\begin{document}
\bibliographystyle{authordate1}

\footnotetext[1]{Universit\'e de Lyon}
\footnotetext[2]{Institut des Syst\`emes Complexes Rh\^one-Alpes (IXXI)}
\footnotetext[3]{Laboratoire de Physique, \'Ecole Normale Sup\'erieure de Lyon et CNRS}
\footnotetext[4]{Groupe d'analyse et de th\'eorie \'economique (GATE), Universit\'e Lyon 2 et CNRS}
\footnotetext[5]{Laboratoire d'\'Economie des Transports, Universit\'e Lyon 2 et CNRS}
\footnotetext[0]{Corresponding author: sebastian.grauwin@ens-lyon.fr}
\maketitle

\newpage
\begin{abstract}
\indent In his 1971's \emph{Dynamic Models of Segregation} paper, the economist Thomas C. Schelling showed that a small preference for one's neighbors to be of the same color could lead to total segregation, even if total segregation does not correspond to individual preferences and to a residential configuration maximizing the collective utility. 

The present work is aimed at deepening the understanding of the properties of dynamic models of segregation based on Schelling's hypotheses. Its main contributions are (i) to offer a comprehensive and up-to-date review of this family of models; (ii) to provide an analytical solution to the most general form of this model under rather general assumptions; to the best of our knowledge, such a solution did not exist so far; (iii) to analyse the effect of two devices aimed at decreasing segregation in such a model.

Chapter one summarizes the ingredients of Schelling's models. We show how the choices of the agent's utility function, of the neighborhood description and of the dynamical rule can impact the outcome of a model. Based on the observation of simulations' results, we find that the neighborhood description does not have a qualitative impact. As regards the dynamical rules, we show that the Logit Behavioral rule introduced in this literature by \cite{young98, zhang2004rsa} presents several advantages relatively to the Best Response rule. 

Chapter two presents a general analytical solution to the model. To that aim, Schelling's model is recasted within the framework of evolutionary game theory, as previously done by \cite{young98, zhang2004rsa}. This allows to define sufficient assumptions regarding agents' utility functions that permit predicting the final state of the system starting from any configuration. This analytical resolution is then used to consider the outcomes of Schelling's utility function and of other utility functions previously used in this context. 

Chapter three examines the effects of introducing coordination in the moving decisions. This coordination is achieved through two different ways. We first impose different levels of taxes proportional to the externality generated by each move of the agents. It is shown that even a low level of tax is sufficient under certain circumstances to significantly reduce segregation. We then investigate the effect of the introduction of a local coordination by vote of co-proprietors, who are defined as the closest neighbors of each agent. It is shown that even a small amount of coordination can break segregation

\vspace{1cm}
\indent J.E.L. classification codes: C63, C72, C73, D62, J15.
\end{abstract}

\newpage
\tableofcontents

\chapter*{Introduction}
\addcontentsline{toc}{chapter}{Introduction}

Ethnic and immigrant residential segregation is a striking feature of most Western cities. Extended measures of segregation have been provided recently for U.S metropolitan areas by \cite{CutlerEtal2008}, \cite{IcelandScopilliti2008} and  \cite{ReardonEtal2008}. 
\cite{CutlerEtal2008} examine a range of potential determinants of immigrant segregation, among which cultural traits of immigrants and xenophobic sentiment among U.S. natives. 
\cite{CardEtal2008} focus on social interactions in whites' preferences. Their results for the 1970-2000 period show evidence of tipping-like behaviors: the rise of the minority share above a certain threshold in a neighborhood leads to a further continuous decrease of the white population. According to these measures, preferences of white families seem to be so that whites' utility in a neighborhood exhibits a sharp decrease beyond a certain minority share. 

This exploration of the temporal paths of minority shares in neighborhoods is based on the prediction of social interaction models that have been developed to understand segregation. 
As soon as $1969$, Schelling proposed a model aiming at formalizing the aggregate consequences of individual preferences regarding the social environment \citep{schelling69, schelling78}. 
The two basic ingredients of Schelling model (1971) are an individual utility function that determines entirely the level of satisfaction enjoyed by an agent in a location and a dynamical rule that drives agents' location changes and therefore the evolution of the city configuration. 
Using an inductive approach, Schelling showed that if the preferences considered are such that an environment of more than $50\%$ of own-group agents is highly preferred to a less than $50\%$ of own-group environment, then the equilibrium configuration exhibits high levels of segregation.
Schelling's 1971 paper (Dynamic models of segregation, \emph{Journal of Mathematical Sociology}) is widely known thanks to this apparently paradoxical effect: 
mild individual preferences for own group neighbors lead to a complete segregation at the global scale.
However, a moment of reflexion suffices to understand that, given the highly asymmetrical utility function, the model could hardly lead to an integrated environment. Later research showed that even a peaked utility function, that is, a function achieving its maximum for a $50\%$- $50\%$ environment, can lead to a fully segregated equilibrium as soon as this function is asymmetric - even in a city where the two groups are equally proportioned \citep{zhang2004rsa,pancs07,barr}.

Criticizing the realism of Schelling's model is straightforward : ignorance of institutional causes of segregation, of income effects, of cities' social structure \ldots Anyway, the model has become a favorite example in the modeling of social systems as the unintended macro-level consequences of individual behavior, and Schelling's 1971 paper is his most widely cited publication (more than 400 as of 2009, May 5th). After years of relatively low citation records, his paper accrues since 2003 around 40 citations per year, showing the renewed interest in his model. It is interesting to notice that citations arise from widely different fields: economics and sociology represent the two strongest contributors (40\% of the total number of citations) but computer science, mathematics and physics gather 24\% of the citations.

This substantial scientific activity has lead to new insights: 
the robustness of Schelling's results towards different definitions of individual's utilities and/or environment \citep{pancs07,Fagiolo2007}; a physical analogue of Schelling's model \citep{vinkovic06}; the interpretation of the emergence of segregation patterns as the result of a coordination problem \citep{zhang2004dmr,zhang2004rsa}. However, these explorations rely almost exclusively on agent-based simulations. 

Some research attempted at solving analytically Schelling's model in order to provide more general results concerning the consequences of individual preferences on segregation levels \citep{mobius2000,pollicott01,zhang2004dmr,zhang2004rsa,dokumaci07}. 

However, to the best of our knowledge, no general analytical framework exists to date that could predict the global pattern emerging from a given utility function. 
Zhang's contribution \citep{zhang2004dmr,zhang2004rsa} represents today the closest achievement in this direction. He proposes a variation of Schelling's model which he analyses formally using a game-theoretical approach and long-range dynamics properties. Unfortunately, his derivation suffers from several deficiencies : 
his calculations are limited to specific utility functions, and even more important, \cite{zhang2004rsa} abandons Schelling's individual rationality. 
Indeed, Zhang's analytical approach cannot deal with vacancies, which leads him to define a new dynamical rule: 
individuals' moves in the city occur when two agents agree on exchanging locations.
This context allows an analytical resolution of the model because it is possible to translate each individual movement into the variation of an aggregate-level function that parallels the segregation level. 

In this paper, we propose a general analytical solution to Schelling's model, i.e. an analytical treatment that allows to calculate the global segregation pattern starting from almost arbitrary individual utility functions. 

Chapter one summarizes the ingredients of Schelling's models. Chapter two presents the general analytical solution, allowing to predict the global state of the system from the knowledge of the individual utility functions. Chapter three examines the effects of introducing coordination in the moving decisions, and shows that even a small amount of coordination can break segregation.\\


\chapter{General formalism}
\section{Basic setup}
\subsection{The city and the agents}
\indent Our artificial city is a two-dimensional $N$x$N$ square lattice with periodic boundary conditions, \emph{ie} a torus containing $N^2$ cells. Each cell corresponds to a dwelling unit, all of equal quality. We suppose that a certain characteristic divides the population of this city in two groups of households that we will refer to as red and green agents. Each location may thus be occupied by a red agent, a green agent, or may be vacant. We denote by $N_V$ the number of vacant cells, and by $N_{R}$ and $N_{G}$ the number of respectively red and green agents. All these numbers are kept fixed over a simulation. The parameter $N$ thus controls the size of the city, the parameter $v=N_V/N^2$ its vacancy rate, and the fraction $n_R=N_R/(N_R+N_G)$ its composition.\\ 
\indent We define a \emph{state} $x$ of the city as a $N^2$-vector, each element of this vector labeling a cell of the $N$x$N$ lattice. Each state $x$ thus represents a specific configuration of the city. We note $X$ the set of all possible configurations, the demographic parameters ($N$, $v$, $n_R$) being fixed.\\ 

\subsection{Neighborhoods}
\begin{figure}[h!]
\begin{center}
\includegraphics[width=8cm]{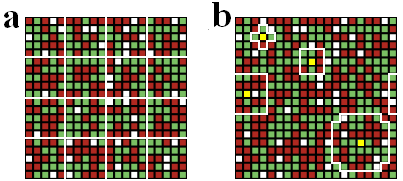}
\end{center}
\caption{\footnotesize{\textbf{Different forms of neighborhood.} Red, green and white squares denote respectively red agents, green agents and vacant cells. \textbf{a.} Example of a bounded neighborhood in which the city is divided in square blocks containing $H+1=25$ cells/locations; \textbf{b.} In the case of a continuous neighborhood description, the neighborhood of an agent corresponds to his $H$ nearest cells/locations. Around the agents marked in yellow, we enlightened by the white frontiers a $H=4$, a $H=8$, a $H=24$ and a $H=44$ continuous neighborhood. [If you printed this document in black and white, the red and green squares should appear respectively in dark grey and soft grey.]}}
\label{vois1}
\end{figure}

\indent Since \cite{schelling69}'s work, two ways of conceiving the neighborhood of an agent have been developed and used in analytical and simulation models.\\
\indent \emph{Bounded neighborhood} models (Fig \ref{vois1}a) describe cities divided into geographical subunits within which all agents are connected. The neighborhood of an agent is thus composed entirely and exclusively of the locations present in the same subunit than his own. In the following, when we refer to a bounded description model, we will implicitly assume that the city is divided into blocks of similar size $H+1$, where $H$ is a fixed integer that corresponds to the number of locations in an agent's neighborhood. Fig \ref{vois1}a displays an example of a city divided into square blocks, which corresponds to the kind of bounded neighborhoods we use in simulations. Bounded neighborhoods are well adapted to models that try to capture or reproduce the effects of the administrative divisions of real cities such as census areas or school districts.\\ 
\indent \emph{Continuous neighborhood} models (Fig \ref{vois1}b) describe cities where the neighborhoods do not correspond to a zoning at the city level, but are centered on the local perception of each agent. In a continuous neighborhood description, one assumes that the neighborhood of an agent is composed of the $H$ nearest locations surrounding him. The $H=4$ ``Von Neumann neighborhood'' and the $H=8$ ``Moore neighborhood'' that are displayed among other examples on Fig \ref{vois1}.b are the most commonly used neighborhoods in agent-based computational models.\\

\indent Note that since some locations remain empty, the size $H$ of the neighborhood of an agent can also be interpreted as the maximum number of neighbors an agent can have. We will discuss in section \ref{NI} the impact of the neighborhood description on the forms of segregation at the city scale. We will see that the global characteristics of our model remain qualitatively independent of any specific definition of neighborhood, provided its size $H$ is relatively small compared to the size $N^2$ of the city, in order to maintain the ``local'' property of neighborhood.  

\subsection{Agent's utility function}
\indent Each agent computes his own level of satisfaction via a utility function which depends only on his 
neighborhood composition. 
let us consider an agent whose neighborhood is composed of $R$ red agents, $G$ green agents and $V$ vacant cells. Since $R+G+V=H$, one needs two independent parameters to describe the composition of the neighborhood of the agent. 
In all generality, we can thus write the utility of an agent for example as a function of $R$ and $G$ or as a function of the fraction $s$ of the agent's similar neighbors and $V$. 
Most models of the literature assume for simplification that agents of a same group share the same utility function. Hence, one only needs a utility function $u_R$ to describe the preference of the red agents and a utility function $u_G$ to describe the preference of the green agents. 
Table \ref{form-ufunc} displays some possible choices of input variables for defining a utility function.\\   
\begin{table}[h!]
\begin{center}
\begin{tabular}{|c|c|c|c|}
\hline
input variables & $(R,G)$ & $(S,V)$ & $(s,V)$\\ 
\hline
utility of a red agent   & $u_R(R,G)$ & $u_R(S=R,V)$ & $u_R(s=R/(R+G),V)$\\
utility of a green agent & $u_G(R,G)$ & $u_G(S=G,V)$ & $u_G(s=G/(R+G),V)$\\
\hline
\end{tabular}
\caption{\footnotesize{Different possible ways of defining the agents' utility functions.}}
\label{form-ufunc}
\end{center}
\end{table}

\indent It is easy to understand that a utility function can be defined up to an additive constant depending on a reference situation but also up to a multiplicative constant depending on the measure scale. 
A common choice in the literature is to take these constants such that a zero utility level denotes a complete dissatisfaction of the agent and a utility of one denote complete satisfaction. We will stick to that use in the following.\\

\begin{figure}[h!]
\begin{center}
\includegraphics[width=14cm]{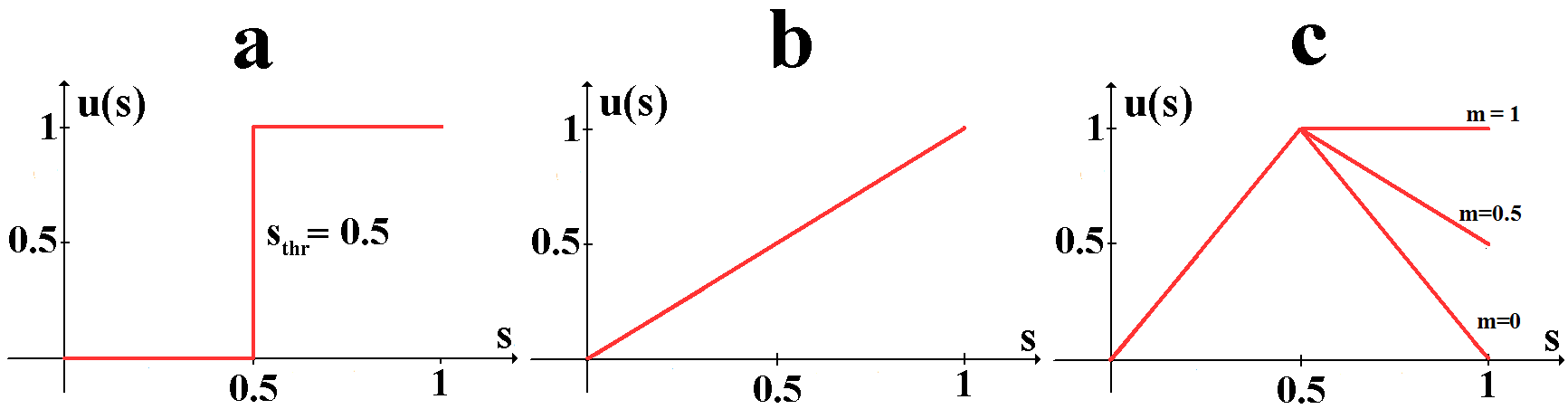}
\end{center}
\caption{\footnotesize{\textbf{Different examples of utility functions presented in the literature.} 
\textbf{a.} Schelling's utility function is a stair-like function for which $u(s)=0$ for $s<s_{thr}$ and $u(s)=1$ for $s\geq s_{thr}$; 
\textbf{b.} the ``linear'' function corresponds to $u(s)=s$ for $0<s<1$; 
\textbf{c.} the ``asymmetrically peaked'' functions correspond to  $u(s)=2s$ for $s<0.5$ and $u(s)=m+2(1-m)s$ for $s\geq 0.5$. They present a maximum for $s=0.5$ but contain an asymmetry in favor of similar neighborhood, this asymmetry being controlled by the parameter $m=u(1)$. For $m=0$, the utility function is symmetrically peaked.}}
\label{util}
\end{figure}

\indent Most models presented in the literature use utility functions that only depend on the sole parameter $s$. Hence they do not take into account the influence of the local vacancy rate, a choice that can be justified \emph{a posteriori} by the fact that the vacant cells are almost always uniformly distributed when the system reach the equilibrium.  
Some of these functions are presented in Figure \ref{util}. 
The most commonly used utility function is Schelling's stair-like function which is equal to one if $s$ is greater than a fixed threshold $s_{thr}$, and equal to zero otherwise (Figure \ref{util}.a). 
However, several critics have been made regarding the artifact generated by these utility functions.
For example, $s_{thr}=0.5$ leads to highly segregated configurations in which the agents are all satisfied, although they would be equally content to live in a $50-50$ neighborhood. 
In this case, while no agent strictly prefers segregation, it is also true that no agent is against it: the apparition of segregation is thus not so surprising. Hence came an urge to investigate less ``binary'' utility functions. 
\cite{bruch06} worked with monotonic increasing functions such as the linear one presented on Figure \ref{util}.b which models agents that are happier as their neighborhood becomes more and more of their own color. 
\cite{zhang2004rsa}, \cite{pancs07} and \cite{barr} worked with functions presenting a maximum for $s=0.5$ such as the asymmetrically peaked ones presented on Fig \ref{util}.c which models agents whose main preference is for mixed neighborhood but who still prefer all-similar neighborhoods to all-dissimilar ones.\\

\indent Few models in the literature present utility functions that depend on two variables. 
\cite{zhang2004dmr} uses utility functions that can be written in terms of $s$ and $V$, where the second variable $V$ is used to give to each location a price that depends on the local density. 
Zhang hence builds a segregative model which incorporates a simple residential market.\\

From a sociological point of view, the origin of the agents' utilities is essential. How do they depend on social policies, history, culture, economy, etc... In this paper, we just accept this individualistic definition of satisfaction and calculate its consequences.  

\subsection{Aggregate measures}
\indent In order to characterize a configuration on the global (city) scale, we need to introduce aggregate measures. 
Let $s_k$, $k\in\{1,..N_R+N_G\}$ be the fraction of agent $k$'s similar neighbors.
In order to characterize the global level of segregation, we introduce for each configuration $x$ the average fraction of same-type neighbors, or \emph{similarity}:
\begin{equation}
\bar{s}(x) = \frac{1}{N_R+N_G} \sum_k s_k 
\end{equation}
Similarity is a well-known measure of segregation that was already used by \cite{schelling71}. 
However, the knowledge of $\bar{s}(x)$ may not be sufficient to determine if the level of segregation of a given configuration is significantly high or low compared to a random configuration with the same demographic parameters: 
a similarity of $0.8$ would point out a high degree of segregation in a city with equally proportioned groups ($n_R=0.5$) but would be insignificant in a city with disproportioned groups ($n_R=0.9$). To avoid this kind of problem, we define in the spirit of \cite{carrington97} the normalized index of similarity $s^*: X\rightarrow [-1,1]$ by
\begin{equation}
s^*(x) =   
\begin{cases}
\dfrac{\bar{s}(x)-\bar{s}_{random}}{1-\bar{s}_{random}}  \mbox{ if } \bar{s}(x) \geq \bar{s}_{random}\\\\
\dfrac{\bar{s}(x)-\bar{s}_{random}}{\bar{s}_{random}}  \mbox{ if } \bar{s}(x) < \bar{s}_{random} 
\end{cases} 
\end{equation}

where $\bar{s}_{random}$ is the expected value of the similarity index $\bar{s}$ implied by a random allocation of the agents in the city. This value depends on the size $N$, vacancy rate $v$ and composition $n_R$ of the city and on the size $H$ of a neighborhood. However, in the cases studied below, i.e. for $n_R=0.5$, $v>0$, $H \geq 1$, $N\gg 1$, the value of $\bar{s}_{random}$ is indistinguishable from $0.5$.\\

\begin{figure}[h!]
\begin{center}
\includegraphics[width=14.3cm]{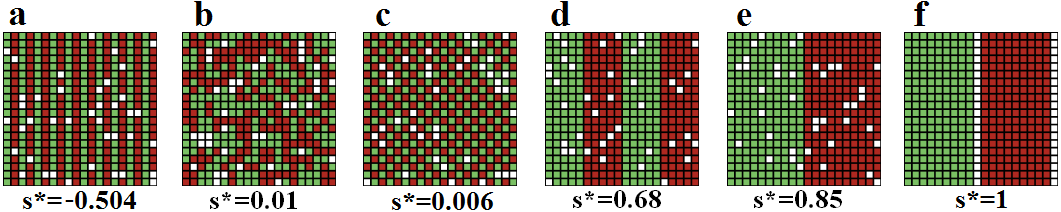}
\end{center}
\caption{\footnotesize{Some examples of configurations of the city, along with their $s^{*}$-values. For ordered configurations such as \textbf{a},\textbf{c},\textbf{d},\textbf{e}, the value of $s^*$ fluctuates with the precise location of the vacant cells. The neighborhood size is $H=8$ and the demographic parameters are fixed to ($N=20$, $v=10\%$, $n_R = 0.5$). }}
\label{s-index}
\end{figure}

\indent Fig \ref{s-index} displays some examples of configurations along with their $s^{*}$-values. 
The reference value $s^{*}=0$ corresponds to an average random configuration. Positive values of $s^{*}$ mean that the agents have more similar neighborhoods than in the random case, and negative values that their neighborhoods contain less similar neighbors than in an average random case. A maximum value $s^*=1$ corresponds to the case where all the neighbors of all agents belong to their own group. Practically, given the random fluctuations of the configurations, an absolute deviation larger than $0.05$ corresponds to configurations which deviate significantly from the random configuration. Notice finally that the normalized similarity index cannot grasp every aspects of the city configurations. Because of the random fluctuations, it can't for example make the difference between a `checkerboard' (Figure \ref{s-index} c) ordered configuration and a random configuration (Figure \ref{s-index} b) \footnote{For a complete and detailed discussion on segregation indices (which are used, what properties should they verify, etc...), see for example \cite{massey88}, \cite{reardon02}, \cite{reardon04}.}.\\

\indent We also introduce notations in order to characterize the level of collective utility: 
\begin{eqnarray}
U(x)&=&\sum_k u_k\\
U^*(x)&=& \frac{1}{N_R+N_G} U(x) \\\nonumber
\end{eqnarray}
where $u_k$ is the utility of agent $k$, $U(x)$ denote the collective utility of a configuration $x$ and $U^*(x)$ its normalized value. 


\section{Dynamic rules}
\indent The core of the kind of model we are dealing with is that agents are given opportunities to move to increase their individual utilities. 
Once the static description of the model is specified, one must add a dynamic rule that governs these moves. 
In the following, we will impose that:
\begin{itemize}
\item At the first iteration, an initial configuration is randomly chosen.
\item At each iteration, one agent and one vacant cell are picked at random.
\item The picked agent then chooses to move in that vacant cell with a probability $Pr\{move\}$ that depends on the utility gain $\Delta u$ he would achieve if he was to move, this probability still being to be specified. 
\end{itemize}

\indent Instead of assuming, as Schelling did, that the agents move to the nearest satisfactory position (the idea being that the cost of moving increases with distance), we suppose here that the distance between the current and envisaged locations of an agent does not intervene in his deciding whether to move or not\footnote{This could be justified by assuming that the cost of moving for the largest possible distance is smaller than any possible strictly positive difference in utility between two locations.}.\\
\indent We present in this section several possible ways of defining the probability $Pr\{move\}$ which are adapted from what exists in the literature. 
Our goal is to discuss their respective degrees of realism but also their respective efficiency: 
we will see that the choice of the probability $Pr\{move\}$ can strongly change the analytical properties of a model, its sensibility to the initial configuration or the nature of its final configurations.

\subsection{Myopic best-response}
\indent Given the preferences, the behavioral assumption made by Schelling (1969, 1971, 1978) is that of myopic best-responses (BR): 
an agent will decide to move if and only if it increases his utility. 
Furthermore, the agents are supposed to have some inertia, which means that an agent moves only to strictly increase his utility. 
The probability that the picked agent chooses to move can thus be written as:
\begin{eqnarray*}
Pr\{move\}&=1 &\mbox{if $\Delta u >0$}\\
Pr\{move\}&=0& \mbox{otherwise}
\end{eqnarray*}
\indent A variation on this kind of dynamic is to remove the inertia and to suppose that an agent may sometimes take utility-neutral moves. 
The probability $Pr\{move\}$ could in this case be written as 
\begin{eqnarray*}
Pr\{move\}&=1 &\mbox{if $\Delta u > 0$}\\
Pr\{move\}&=0.5& \mbox{if $\Delta u = 0$}\\
Pr\{move\}&=0& \mbox{otherwise}
\end{eqnarray*}
In the following, we refer to these two kind of dynamic rules respectively as `strict BR' and `non-strict BR'.\\
\indent In a study bearing on different utility functions, \cite{pancs07} show that strict BR often leads to steady-states where no agent can find a vacant cell which would strictly improve its utility, even if a large fraction of agents are still unsatisfied. 
The artificial city hence quite often ends up being stuck in a ``frozen'' configuration. 
These steady-states hence constitute Myopic Nash Equilibria (MNE), the term myopic being justified by the fact that the unsatisfied stuck agents disregard the fact that taking a utility-neutral or a utility-decreasing move at a given moment may allow them to reach later a better position than their current one. 
This dynamic is neither realistic (since people include other criteria to decide whether moving or not) nor analytically convenient, since the appearance of frozen configurations renders the final states of the system dependent on the initial states (see below, Figs \ref{dyn-usch} and \ref{dyn-uap}). Furthermore, \cite{singh2007} show that strict BR dynamics leads to scale-dependent results because of blocking. With stair-like individual utility functions, they obtain complete segregation for a small city, but only partial segregation as the size of the city increases.\\ 
\indent \cite{vinkovic06} interpret the removal of the inertia as the introduction of some `fluidity' in the system. 
In practice, most of the inconvenience of the strict BR dynamic disappear by removing the inertia, the system being then able to get out of most of the frozen configurations. Non-strict BR dynamics are then less dependent of the initial states. 
But Myopic Nash Equilibria, although less numerous under these hypotheses, still exist and in theory the final outcomes may still depend on the initial state. 

\subsection{A better behavioral rule: the logit dynamical rule}\label{abbr}
\indent The probability to move that we present here goes a little bit further than non-strict BR by allowing utility-decreasing moves. We assume that the picked agent chooses to move to the picked vacant cell with a probability:
\begin{equation}
Pr\{move\} = \frac{1}{1+e^{\,-\Delta u/T}} 
\label{logit}
\end{equation}
where $T>0$ is a fixed parameter.

\indent The scalar $T$ can be interpreted as a measure of the level of noise in an agent's decision. 
Clearly, the probability for an agent to take a utility-decreasing move drops down as $T \rightarrow 0$ and the described rule thus converges to the non-strict BR rule. 
For any finite $T>0$, the agents choose non-best replies with a non-zero probability, but actions that yield smaller payoffs are chosen with smaller probability.\\
\indent  This kind of perturbed best-response dynamics has been developed in e.g. \cite{anderson92} or \cite{young98}. 
Taken as a behavioral rule, the underlying logit choice function in eq. \ref{logit} is rooted in the psychology literature \citep{thurstone1928}. 
From the microeconomic point of view, it can be given a justification in terms of a random-utility model (see appendix \ref{Alogit}) where the random part in the utility function can be interpreted as a way to take into account criteria other than the neighborhood composition such as the quality of the housing, the proximity to one's workplace or any other idiosyncratic amenity.\\

\indent Beside being more realistic from a behavioral point of view, the logit rule also provides a strong analytical framework to Schelling's model.
Obviously, it implies that the probability that the state at the $t^{th}$ iteration $x^{t}$ is equal to a given state $x$ only depends on the state at the previous iteration $x^{t-1}$: 
\begin{equation}
\Pr(x^{t}=x|x^{t-1}, \ldots, x^1, x^0) = \Pr(x^{t}=x|x^{t-1})
\end{equation}
The dynamic rule thus yields a finite Markov process.\\
\indent It is then easy to figure out that the Markov chain describing our system is irreducible (since $T>0$ each imaginable move has a non-zero probability to happen and it is thus possible to get to any state from any state), aperiodic (given any state $x$ and any integer $k$, there is a non-zero probability that we return to state $x$ in a multiple of $k$ iterations) and recurrent (given that we start in state $x$, the probability that we will never return to $x$ is $0$). These three properties ensure that the probability to observe any state $x$ after $t$ iterations starting from a state $y$ converges toward a fixed limit independent of the starting state $y$ as $t\rightarrow \infty$.\\
\indent In other words, for each set of parameters and dynamic rule, there exists a stationary distribution 
\begin{equation}
\Pi: x \in X \rightarrow \Pi(x) \in [0,\,1]\,\,, \sum_{x\in X}\Pi(x) = 1
\end{equation}
which gives the probability with which each state $x$ will be observed in the long run.\\
\indent Clearly, for $T\rightarrow \infty$, the randomness introduced in the dynamical rule prevails and the stationary distribution is just a constant. Similarly, for any finite $T>0$, our dynamical system (the city) is evolving toward an attractor composed of a subset $A$ of $X$. It follows that any measure $\mathcal{M}$ performed on the states space $X$ - such as the global utility $U$ - will in the long run fluctuate around a mean value $\mathcal{M}_{\infty} = \sum_{x\in A} \Pi(x)\mathcal{M}(x)$. 
These mean values may depend on the amplitude of the noise $T$, but the amplitude of the fluctuations decreases as $T\rightarrow 0$. These intuitions are confirmed later.\\

\indent In the following, we refer to two states $x$ and $y$ as \emph{immediately communicating states} (ICS) if we can switch from the state $x$ to the state $y$ by moving one single agent. We also note $\Delta_{xy}u$ the variation of the utility of that agent induced by this particular move and $P_{xy}^T$ the probability to be in state $y$ at a given iteration if the system was in state $x$ at the previous iteration. 
According to the dynamic rule presented in the previous section, one has:
\begin{eqnarray}
P_{xy}^T &=\,\, \gamma (1+e^{\,-\Delta_{xy} u/T})^{-1} &\mbox{if $x$ and $y$ are ICS} \label{trans} \\
P_{xy}^T &=\,\, 0 &\mbox{if $x$ and $y$ are not ICS} 
\end{eqnarray}
where the parameter $\gamma=1/(N_V(N_R+N_G))=1/(v(1-v)N^4)$ takes into account the probability to pick the right agent and the right vacant cell that allow to pass from $x$ to $y$. $P^T$ thus corresponds to the probability transition matrix for a fixed $T$ and the stationary distribution $\Pi$ is by definition the unique normalized function defined on $X$ that verifies for all $x\in X$
\begin{equation}
\sum_y P_{yx}^T\Pi(y) = \Pi(x)
\end{equation}

\indent let us insist on the the fact that, compared to the BR rules, the final states of the system don't depend here on the initial ones (which is,   in terms of a model's effectiveness, highly convenient). The simulations we present in the next section fortunately show that the outcomes generated by the BR and logit dynamic rule are in general rather close. 
This is understandable, since we already pointed out that the limit $T\rightarrow 0$ corresponds to a non-strict BR dynamic rule. 
For low values of the noise $T$, a configuration corresponding to a MNE in the BR dynamic should then have a relatively high probability to appear when using the logit rule.
This intuition can be confirmed by the following property: 
\begin{center}
\begin{flushleft} \hspace{0.5cm} \underline{Lemma 1}\\\end{flushleft}
\fcolorbox{black}{gris5}{
\begin{minipage}{1\textwidth}
Let $t_{xy}$ be the expected time for the system leaving the state $x$ to reach the state $y$.\\
In the case when the stationary distribution $\Pi$ is unique and well defined, the stationary probability to observe a state $x$ is related to the expected return time $t_{xx}$:
$$\Pi(x) = 1/t_{xx}$$
\end{minipage} }
\end{center}

\indent A proof of this lemma is reproduced in appendix \ref{Alemma1}. 
It is pretty obvious that if $x$ is a configuration corresponding to a MNE in the BR dynamic, in the limit $T\rightarrow 0$, the expected return time $t_{xx}$ converges towards low values. Lemma 1 thus ensures that the probability to observe $x$ in the long run is rather high.\\


\section{Simulations}
\indent In order to get a better grasp of how a model following our prescriptions behaves, we present in this section some simulation results.\\

\subsection{Impact of the neighborhood description}
\label{NI}
\begin{figure}[h!]
\begin{center}
\includegraphics[width=14cm]{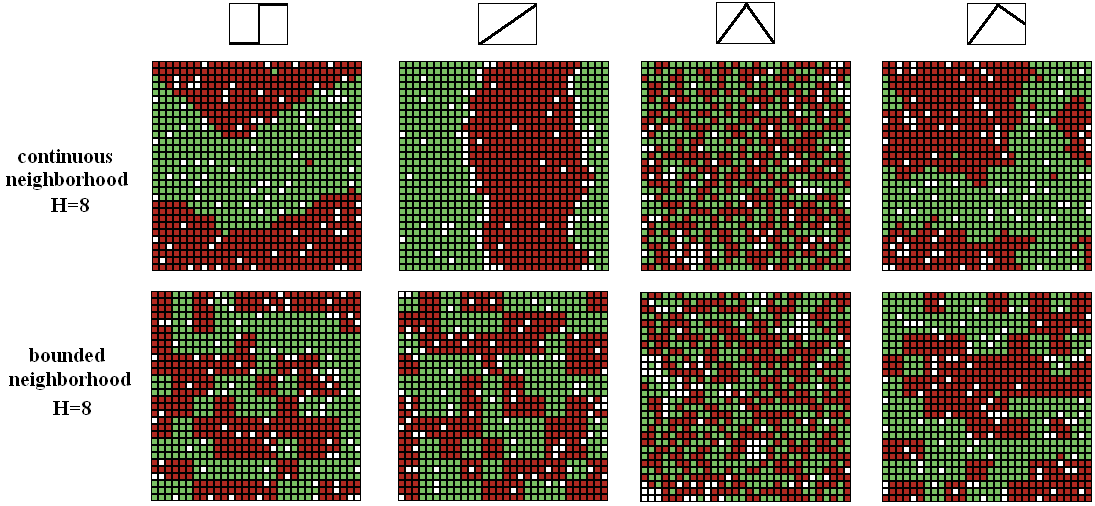}
\end{center}
\caption{\footnotesize{\textbf{Typical stationary configurations obtained by simulations.} The demographic parameters are $(N=30,v=10\%,n_R=0.5)$. Neighborhood size is fixed to $H=8$. \textbf{Top panel.} With a continuous neighborhood description. \textbf{Bottom panel.} With a bounded neighborhood description. From left to right: the agents compute their utility with some of the functions presented earlier (the stair-like function with $s_{thr}=0.5$, the linear function, the symmetrically peaked function and the asymmetrically peaked function with $m=0.5$, all functions figured as pictograms on the top row). The simulation were run using the logit dynamical rule with a level of noise $T=0.1$.}}
\label{influNu}
\end{figure}

Let us consider the snapshots presented on Figure \ref{influNu}. They present typical final states which have been obtained by running simulations using the logit dynamical rule and different utility functions and which allow to compare results depending on the chosen neighborhood description. 
It is noticeable that when global segregative patterns appear in the continuous description, segregative patterns following block boundaries -``checkering''- appear in the bounded description. Since there is no interaction between agents of different blocks, there is no reason why blocks of same color should aggregate, and checkering thus corresponds in the bounded description to the pattern of maximal segregation. In the same way, when simulations are run with the symmetrically peaked utility function, mixed random configurations are obtained in both bounded and continuous neighborhood description. 
On the basis on these few observations, one can conjecture that the choice of a neighborhood description is qualitatively not determinant on the outcome of the models regarding the segregation issues. 
In appendix \ref{Aneigh}, we present other simulation results which tend to prove that, when using the logit dynamic rule, the outcome of the models are also qualitatively independent of the neighborhood size $H$.\\

\subsection{Welfare \emph{versus} segregation issues}
\indent The snapshots on Figure \ref{influNu} also allow to compare results according to the chosen utility function of the agents (for simplicity, we choose here cases where red and green agents share the same utility function $u(s)$). 
In the cases of the stair-like and of the linear utility functions, for which the agents have a preference for like-neighbors, large segregation patterns appear. Note that while the utilities of almost all the agents are maximized within these configurations, the fact that large segregation patterns appear in the case of the stair-like function is not trivial: a lot of configurations presenting only local segregation patterns also maximize the collective utility. In the case of the ``symmetrically peaked'' utility function, the system converges toward randomly-organized mixed configurations which also maximize the utility of most agents. Finally, the case of the asymmetrically peaked function may be the most intriguing: large segregative patterns appear although they absolutely do not maximize the utility of most agents (most of them being stuck inside an homogeneous area with a utility of $0.5$).\\
\indent In this case, one of the key element driving the segregation is the asymmetry of this utility function, ie, the fact that even if the agents have a strict preference for mixity, they still favor a large-majority status over a small-minority status.  Another key element is the fact that the agents take only selfish decisions. Actually, as argued by \cite{zhang2004rsa} and \cite{pancs07}, the reason why individual preferences for integrated environments may lead to segregated configurations can be expressed in economic terms as the existence of externalities. Indeed, location choice by an agent is only based on her own utility level, even if it also affects her neighbors' utility levels. 
In particular, with the asymmetrically peaked utility function, a red (green) agent may move for example from a $49\%$ red (green) neighborhood to a $51\%$  red (green) neighborhood because it slightly increases her utility. 
Meanwhile, this move is likely to decrease the utility of the agents in the initial and final moving agent's neighborhoods and therefore decrease the collective utility level. 
Both of these factors imply that a highly-segregated configuration is necessarily very stable. 
Indeed, once the city is divided into homogeneous areas, a red agent will have no incentive to go from the red area to the green one (his utility dropping from $0.5$ to $0$)\footnote{
And even though a red agent goes from time to time into the green area by mistake, because of the asymmetry in the utility function, he will have a strong incentive to return to the red area, which he will do very likely before a second red agent rejoins him in the green area.}.\\

\begin{figure}
\begin{center}
\includegraphics[width=7cm]{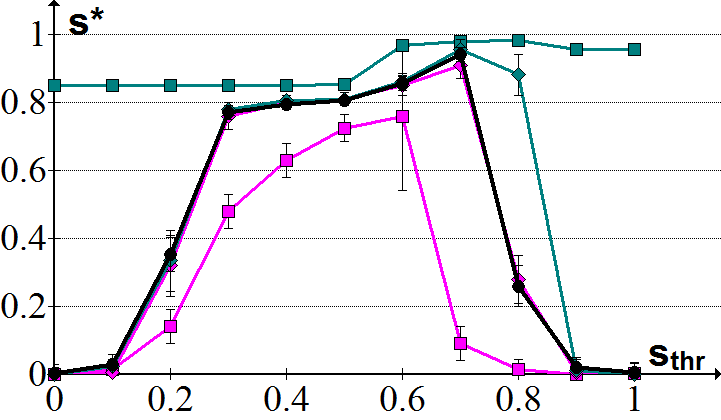}\hfill
\includegraphics[width=7cm]{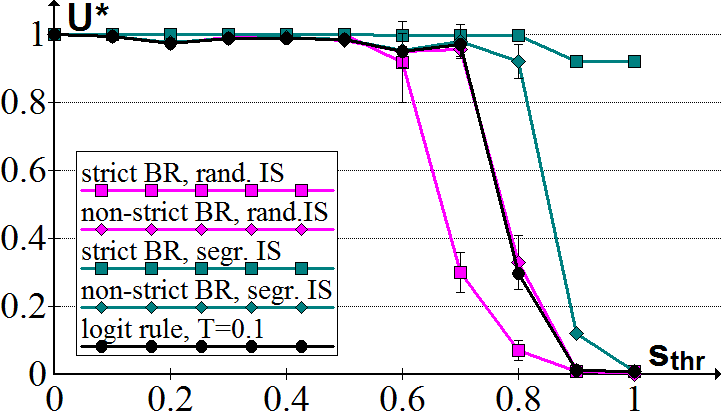}
\end{center}
\caption{\footnotesize{\textbf{Stair-like utility function.} Stationary values of the normalized similarity index and of the normalized collective utility obtained by simulating the move of agents whose utility is computed using a stair-like function u(s) with a threshold $s_{thr}$ varying from 0 to 1. We used a $H=8$ continuous neighborhood description and the demographic parameters were fixed to $(N=30, v=10\%, n_R=0.5)$. Simulations are started from random or segregated initial states.}}
\label{dyn-usch}
\end{figure}

\subsection{Dynamical issues}
\label{dyn-issues}
\begin{figure}
\begin{center}
\includegraphics[width=7cm]{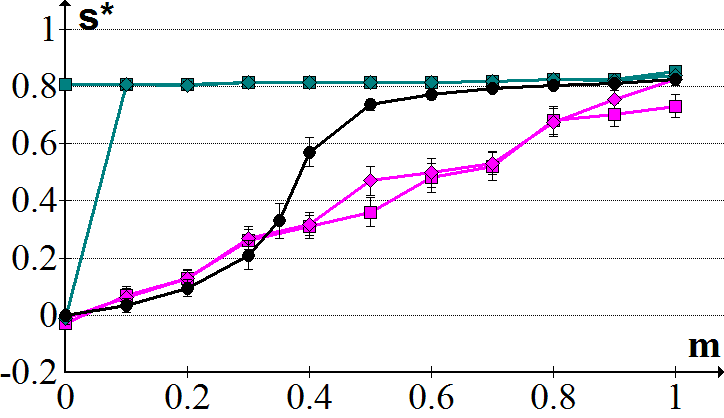}\hfill
\includegraphics[width=7cm]{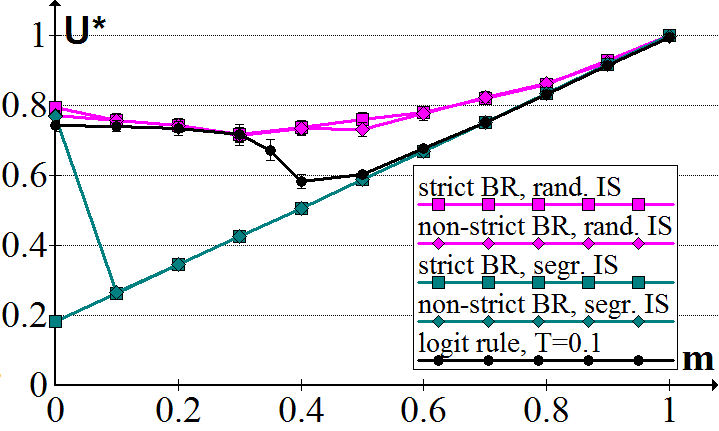}
\end{center}
\caption{\footnotesize{\textbf{Asymmetrically peaked utility function.} Stationary values of the normalized similarity index and of the normalized collective utility obtained by simulating the move of agents whose utility is computed using a asymmetrically peaked function $u(s)$ parametrized by $m\equiv u(s=1)$ varying from 0 to 1. We used a $H=8$ continuous neighborhood description and the demographic parameters were fixed to $(N=30, v=10\%, n_R=0.5)$. Simulations are started from random or segregated initial states.}}
\label{dyn-uap}
\end{figure}

\begin{figure}
\begin{center}
\includegraphics[width=6.5cm]{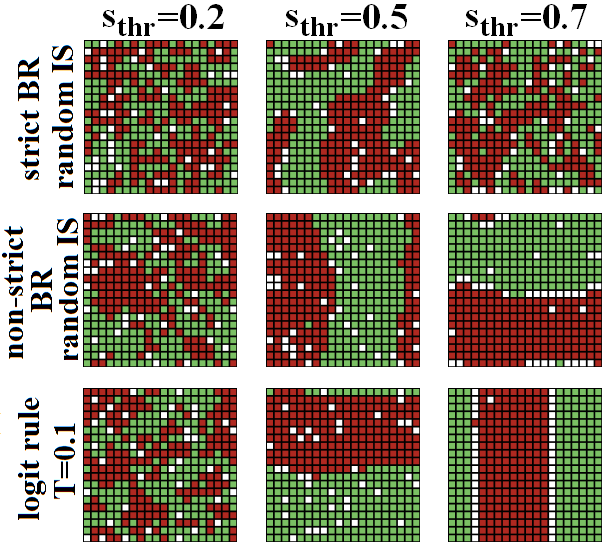}\hfill
\includegraphics[width=6.5cm]{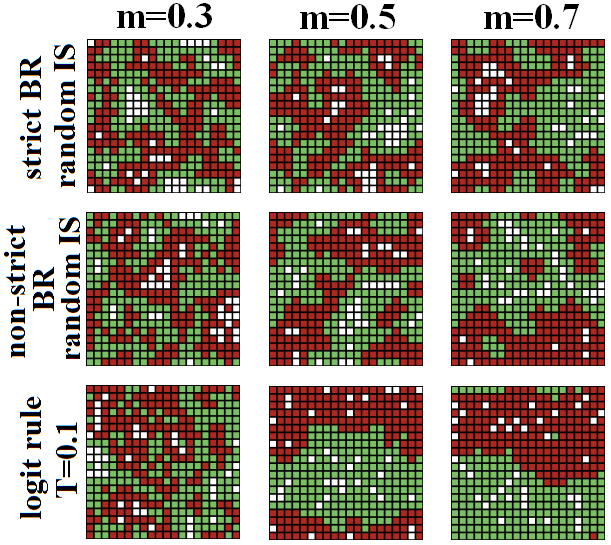}\hfill
\end{center}
\caption{\footnotesize{\textbf{Left.} Snapshots of stationary configurations obtained using stair-like utility functions, which correspond to some of the results presented in fig \protect \ref{dyn-usch}. \textbf{Right.} Snapshots of stationary configurations obtained using asymmetrically peaked utility functions, which correspond to some of the results presented in fig \protect \ref{dyn-uap}. }}
\label{snapshots}
\end{figure}

\indent Figures \ref{dyn-usch} and \ref{dyn-uap} display stationary values of the normalized similarity index and of the normalized collective utility obtained by simulating the move of agents whose utility is computed using respectively a stair-like function and an asymmetrically peaked function. Snapshots of some of these stationary configurations are presented in Figure \ref{snapshots}. 
To contrast the three dynamic rules, these simulations have been performed using alternatively the strict BR, the non-strict BR and the logit dynamic rule. Figures \ref{dyn-usch} and \ref{dyn-uap} clearly show that the two best-response rules lead to a strong dependence of the final state on the initial configuration\footnote{More precisely, in the case of a stair-like utility function, non-strict BR dynamics generally leads to final configurations that do not depend on the initial configuration, as shown by \cite{vinkovic06}. However, using other utility functions, such as the asymmetrically-peaked one, cancels this effect (Figure \ref{dyn-uap}), and one needs the logit rule to break the dependence on the initial state.}.\\
\indent For each simulation, we start both from a randomly chosen initial state or a totally segregated initial state similar to the ones presented on fig \ref{s-index}b and e. In the case of the BR dynamics, we wait until the system gets caught in a frozen state or until each agent is given in average $1000$ opportunities to move; we then repeated this procedure $200$ times for each set of parameters. Each corresponding point and error bar in figures \ref{dyn-usch} and \ref{dyn-uap} are the mean values and the standard deviation of $s^*$ or $U^*$ over these $200$ procedures. In the case of the logit dynamics, we start from random and totally segregated configurations and wait until both initial states converge to similar final configurations (i.e. similar $s^*$ or $U^*$ values). In this case, error bars represent standard deviations in the final values, which correspond to arbitrarily stopped simulations.\\   

\subsubsection*{Stair-like utility function}
\indent In the case of the stair-like utility function, low values of $s_{thr}$ correspond to situations where individuals are satisfied over a large range of neighborhood composition: 
as soon as the share of same-color neighbors is above $s_{thr}$, the maximum utility level is reached. 
Accordingly, figure \ref{dyn-usch} shows that the normalized similarity index $s^*$, which represents the average fraction of same-type neighbors, is small for values of the threshold lower than $0.2$.
It increases however quite sharply when the threshold changes from $0.2$ to $0.3$. 
Consequently and as could be expected, the collective utility level is at its maximum for $0 \leq s_{thr} \leq 0.5$: in these cases and starting from a random configuration, it is quite easy for an agent to find a neighborhood with at least $s_{thr}$ neighbors.
By this move, this agent is not very likely to decrease his neighbors' utility level, as the chance of them seeing the share of like neighbors to decrease below $s_{thr}$ is low.  
For large values of $s_{thr}$, the index $s^*$ decreases dramatically, which can be easily explained by the fact that 
agents prefer segregated local environments only if they are highly segregated. 
If they can not find a neighborhood with a share of same-type neighbors above $s_{thr}$, the agents are indifferent about which location to choose. 
This is a kind of coordination problem, as finding such environments starting from a random configuration is not easy, although it is collectively desirable. 
Contrarily, it seems easier when the threshold $s_{thr}$ is around 0.5 to find ``by chance'' neighborhoods which are slightly above this value. 
In these cases, agents are both willing and able to find segregated neighborhoods. 
This explains the rise in the segregation index, which is at its highest for $s_{thr}$ equal to 0.6 or 0.7.
As far as the dynamic rule is concerned, it is worth noting that the removal of the inertia, compared to the BR rule, increases the segregation level in all cases.
By letting the agents make mistakes, no-inertia rules allow to achieve higher segregation levels.
In particular, for $s_{thr}=0.7$, allowing moves which do not strictly improve the segregation experienced by the agents permits the system to find configurations where the collective utility is higher because the average segregation level is far above the desired threshold.

\subsubsection*{Asymmetrically peaked utility function}
\indent The asymmetrically peaked utility function represents situations in which the agents prefer mixed neighborhoods (with a strict preference as long as $m < 1$).
Figure \ref{dyn-uap} shows a continuous increase in average segregation with the rise of $m$: the stronger the asymmetry, the higher the incentive to move from location below-$50\%$ of like-neighbors to locations above-$50\%$ of like-neighbors. 
The collective utility level first decreases with a rise in $m$ from 0 to 0.3: due to the increase in $s^*$, the agents are less happy because they individually prefer perfectly mixed neighborhoods.
With the BR dynamic rules, collective utility rises with $m$ if it is above 0.3: the higher $m$, the lower the decrease in utility in segregated neighborhoods compared to perfectly mixed neighborhoods. 
However, adopting the logit rule with $T=0.1$ leads to a jump in the average segregation level for $m$ going from 0.35 to 0.5. 
Here as in the stair-like utility function, allowing agents to make some mistakes leads to form segregated neighborhoods, which will afterwards be strictly preferred over environments where the agents are in minority. 
However, this formation of segregated configurations is accompanied by a decrease in the average utility: for medium values of $m$, mixed configurations are still collectively much more preferable to segregated ones. 
Collective utility increases when $m$ increases further, because agents, although they prefer mixed neighborhoods, favor then the majority status over the minority status.
At the end, when $m=1$, the agents are indifferent between mixed environments and environments with $100\%$ of same-type neighbors. 
The outcome in this case is a perfectly segregated configuration that allow agents to achieve the highest utility level.  


\section{Discussion}
\indent An important goal of Schelling-type models is to provide results concerning the determinants of segregation. 
Such results are useful to derive some welfare and policy implications. 
Results of the previous section give a large overview of the outcomes of the model depending on a great variety of parameters (form of the utility function, values of the parameters governing this function, dynamic rule). 
However, one can still wonder which are the general conditions yielding to segregation. 
Suppose in particular that there is a benevolent planner able to affect the agents' preferences and that she aims at maximizing integration.
What preferences would she like them to have? 
In other words, what would be the kind of preferences, if any, that could avoid the segregative outcome? \\
\indent To answer this question, it is worth relying not only on simulations, but also on analytical results. 
To date, the existing analytical results of Schelling-type models are based on specific hypotheses, as argued in the introduction of this paper.
Our aim in Chapter 2 is to provide a generic analytical ground to a Schelling-type model, that will enable us to tell, for a broad family of utility functions, which configurations exist at equilibrium. \\
\indent Furthermore, the gap between micro-motives and macro-behaviors underlined by Schelling is the result of externalities: 
an agent who moves according only to his own interest creates externalities since his move also affects the utilities of the agents living in the neighborhood he leaves and in the neighborhood he arrives in.
Classically, in economics, one can think of introducing some kind of coordination between the agents to avoid the consequences of these externalities, 
either by introducing a coordination mechanism between the agents, or by imposing a taxation on individual moves. 
\cite{pancs07} consider the latter solutions as follows:
``\emph{[...] presuming  that  there  is  a  social  welfare  case  for  integration  (independent  of  the specification of the individual preferences), could a migration subsidy or tax system prevent segregation and implement integration? Although we do not explicitly analyze this issue, our analysis suggests that a system consisting of rewards for integrating moves or taxation of segregating moves might not work if it merely emulates the incentive structure represented by the various utility functions analyzed in our paper.}''\\
\indent We analyze in chapter 3 these two kinds of coordination devices by means of simulations. We will see that in certain cases, introducing coordination is actually sufficient to change the basic model outcomes. 


\chapter{An analytically soluble model}
\section{Introduction}
\indent Schelling's model was first recast within the framework of evolutionary game theory by \cite{young98} who analyzes a one-dimensional Schelling model.
Later, \cite{zhang2004dmr} prolonged this approach by analytically solving a two-dimensional segregation model in which one category of agents is color-neutral while the other category has a utility function which is always increasing in the number of same-color agents\footnote{
\cite{zhang2004dmr} also considers an asymmetric peaked utility function but does not solve this case analytically.}.
\cite{zhang2004rsa} studies the case of an asymmetric peaked utility function within the same framework.
However, \cite{zhang2004rsa} relies on a particular context with no vacant locations, in which individuals' moves in the city occur when two agents agree on exchanging locations. 
This context allows an analytical resolution of the model because each individual movement can be translated into the variation of an aggregate-level function, called the potential function, that parallels the segregation level. 
As stated by the author, ``[the] \emph{kind of externality from social dynamics is made transparent in the potential function of our model that relates segregation, a non-optimum social outcome, to utility gains at the individual level.}''
However, the absence of vacant locations in \cite{zhang2004rsa} implies that agents' moves only occur when two agents agree to do so, that is, if the sum of their utilities is positive. 
This condition is somewhat restrictive and implies some kind of coordination between the agents, which is not present in the basic Schelling's model. 

\indent The present chapter goes a step further by proposing an analytical resolution of a more general model with vacant cells and with general utility functions. We show that it is possible, in the context of a bounded neighborhood description, to define a large class of utility functions for which each individual location choice can be translated into the variation of an aggregate function that characterizes the segregation level in the city. To establish this critical link between the individual move and the global characteristics of our system, our key trick is to use bounded neighborhoods, which allows to reduce agents' heterogeneity by giving the same utility to all inhabitants sharing the same color and the same bounded neighborhood. This `collectivisation' of the individuals prevents any loss of information on the global configuration change, allowing to keep track on how each individual move affects the global configuration. Instead, when continuous neighborhoods are used, this information is lost because the way a moving agent affects his past and new neighbors depends on factors (their neighbors' neighbors) that she does not take into account when moving. Therefore, it is generally impossible to know how an individual move affects a function of the global configuration. Using bounded neighborhoods provides therefore a very general solution, thanks to which it is possible to predict the long-term configurations of the city. Although the potential function is not generally linked with the segregation level, the analysis in term of potential function yields criteria allowing to predict the consequences of very different utility functions in terms of segregation or integration.\\
\indent As an example, Schelling's original stair-like utility function leads to the well-known Duncan dissimilarity index. The potential functions corresponding to other well-known examples of utility functions are also derived. 


\section{Framework}
We place ourselves in a bounded neighborhood framework. The city is divided into a set $\mathcal{Q}$ of blocks, each of which contains $H+1$ locations (hence, the relation $|\mathcal{Q}|(H+1)=N^2$ must hold). For a given configuration $x\in X$ of the city, we denote by $R_q(x)$ and $G_q(x)$ the number of red and green agents that live inside the block $q \in \mathcal{Q}$. Taking into account that some locations of each block may remain empty, the $\{R_q\}$ and the $\{G_q\}$ must thus verify :
\begin{eqnarray}
&\sum_q R_q = N_R&\\
&\sum_q G_q = N_G&\\
&(R_q,G_q) \in E_{H+1}\equiv \{(R,G), 0\leq R+G\leq H+1\} & \label{block-constraint}
\end{eqnarray}

\indent Without any loss of generality, we write the utility of an agent as:
\begin{eqnarray*}
u=u_R(R_q-1,G_q) &\mbox{for a red agent living in block $q$} \\
u=u_G(R_q,G_q-1) &\mbox{for a green agent living in block $q$}
\end{eqnarray*} 
The utility of an agent is thus a function of $E_H\rightarrow \mathbb{R}$. Note that the choice of the input variables of the $u_R$ and $u_G$ functions is consistent with the fact that an agent living in the block $q$ will have $R_q-1$ red neighbors and $G_q$ green neighbors if he is red and similarly $R_q$ red neighbors and $G_q-1$ green neighbors if he is green.


\section{Potential function}
\subsection{Definitions and properties}\label{def-prop}
\indent In game theory, a game is said to be a potential game if the incentive of all players to choose their strategy can be expressed in one global function, which is called the potential function. 
Games can be either ordinal or cardinal potential games. 
In cardinal games, the difference in individual payoffs for each player from individually choosing one's strategy \emph{ceteris paribus} has to have the same value as the difference in values for the potential function. 
In ordinal games, only the signs of the differences have to be the same.\\ 
\indent The concept of potential function was proposed by \cite{monderer96} in which more formal definitions can be found. 
In our context, the definition of a potential function takes the rather simple following form:\\

\indent \underline{Definition}\\
\indent Let $\mathcal{F}:x \in X \rightarrow \mathcal{F}(x) \in \mathbb{R}$ be an aggregate function describing each of the possible configurations.
By definition, $\mathcal{F}$ will be a (cardinal) potential function of our model if and only if each gain in utility $\Delta u$ of a moving agent is equal to the variation $\Delta\mathcal{F}$ that is induced at the global level by the move of this agent.
A cardinal potential function will thus verify: $\mathcal{F}(y)-\mathcal{F}(x)=\Delta_{xy} u$ with $\Delta_{xy} u$ previously defined (section \ref{abbr}).\\

\indent In our model, the main property of a potential function is to link the variation of a purely individual function (the utility of the moving agent) to the variation of a global function defined on the space $X$ of all possible configurations. The fact that the knowledge of $\Delta u$ is sufficient to say something at the global level is highly non-trivial since (for example) there is no way to determine the externalities produced by the move of an agent - \emph{ie} the variation of the utility of his former and new neighbors - only by the knowledge of $\Delta u$. The ensuing lemma points out even more the value of the potential function as an analytical tool.\\ 

\begin{center}
\begin{flushleft} \hspace{0.5cm} \underline{Lemma 2}\\\end{flushleft}
\fcolorbox{black}{gris5}{
\begin{minipage}{1\textwidth}
If $\mathcal{F}$ is a potential function of the system, then the stationary distribution $\Pi$ for any configuration $x$ can be written as
\begin{equation}
\Pi(x)=\frac{e^{\mathcal{F}(x)/T}}{\sum_{z\in X} e^{\mathcal{F}(z)/T}} \label{PPi}
\end{equation} 
It follows that for $T\rightarrow 0$, the stationary configurations are those which maximize $\mathcal{F}$.\\
\end{minipage} }
\end{center}

\indent The following proof follows the classical argument presented in \cite{young98}.\\

\indent Let $\pi$ be the function defined as $\pi: X \rightarrow [0,\,1];\, x\rightarrow\,\pi(x)= e^{\mathcal{F}(x)/T}/\sum_z e^{\mathcal{F}(z)/T}$. The first step of the proof consists in checking that $\pi$ satisfies the \emph{detailed balance condition}: 
\begin{equation}
\pi(x)P_{xy}^T=\pi(y)P_{yx}^T
\label{db}
\end{equation} 
If $x$ and $y$ are two different and not communicating states, the relation \ref{db} is trivially satisfied since in this case $P_{xy}^T=P_{yx}^T=0$. If $x=y$, the detailed balance condition is also trivially verified. In the case where $x\neq y$ and $x$ and $y$ are two communicating states, one has:
\begin{eqnarray}
\pi(x)P_{xy}^T &=& \pi(x)\gamma\frac{1}{1+e^{-\Delta_{xy} u/T}} \,\,=\,\, \pi(x)\gamma\frac{1}{1+e^{-(\mathcal{F}(y)-\mathcal{F}(x))/T}} \,\,=\,\, \pi(x)\gamma\frac{e^{\mathcal{F}(y)/T}}{e^{\mathcal{F}(x)/T}+e^{\mathcal{F}(y)/T}}\nonumber\\
&=& \pi(y)\gamma\frac{e^{\mathcal{F}(x)/T}}{e^{\mathcal{F}(x)/T}+e^{\mathcal{F}(y)/T}} \,\,=\,\, \pi(y)\gamma\frac{1}{1+e^{-(\mathcal{F}(x)-\mathcal{F}(y))/T}} \,\,=\,\, \pi(y)\gamma\frac{1}{1+e^{-\Delta_{yx} u/T}} \nonumber \\ 
&=& \pi(y)P_{yx}^T \nonumber
\end{eqnarray}
recalling that  $\gamma=1/(N_V(N_R+N_G))=1/(v(1-v)N^4)$.

Hence the detailed balance condition is always verified and 
\begin{equation}
\sum_{x\in X}\pi(x)P_{xy}^T=\sum_{x\in X}\pi(y)P_{yx}^T= \pi(y)\sum_{x\in X}P_{yx}^T=\pi(y)\cdot 1=\pi(y)\,,
\end{equation} 
which defines $\pi$ as a stationary distribution of the process. Because the Markov chain is finite and irreducible, it has a unique stationary distribution. Hence, for each state $x$, $\Pi(x) = \pi(x) = e^{\mathcal{F}(x)/T}/\sum_z e^{\mathcal{F}(z)/T}$.\\

\indent Define then $X_F$ as the subset of $X$ of the states that strictly maximize the potential function $\mathcal{F}$: 
\begin{equation}
X_F=\{y,\,\forall x \in X\,\,\, \mathcal{F}(y)\geq \mathcal{F}(x)\}
\end{equation} 
\indent The second part of the lemma can now be proved as follows: for two states $x$ and $y$ of $X_F$, we will have 
$\mathcal{F}(x)=\mathcal{F}(y)$ and therefore $\Pi(x)/\Pi(y) = e^{[\mathcal{F}(x)-\mathcal{F}(y)]/T}=1$,
which means that two states that strictly maximize $\mathcal{F}$ are observed in the long run with the same probability; for two states $x\in X\setminus X_F$ and $y \in X_F$, we will have $\mathcal{F}(x)-\mathcal{F}(y)\leq 0$ and therefore $\displaystyle{\Pi(x)/\Pi(y)= e^{[\mathcal{F}(x)-\mathcal{F}(y)]/T} \rightarrow \, 0}$ as $T\rightarrow 0$. This means that for $T\rightarrow 0$, the probability to observe a state that does not maximize the potential function $\mathcal{F}$ becomes in the long run infinitesimally small $\square$\\
\indent In the case of finite values of the noise level ($T>0$), it can be demonstrated using standard tools of statistical physics that the states which are the more probable to appear are those which maximize $F(x)+TS(x)$ where $S(x)$ is an entropy-like global function taking into account the number of ways of locating $R_q$ red agents and $G_q$ green agents in each block $q$ of the city (refer to \cite{econophys-unpublished} for more precision). \\

\indent The potential function is hence a very powerful analytical tool. 
By establishing a relation between individual changes in utility and a global characteristic of the city configuration, and because stationary configurations can be defined as those maximizing the potential function, this analysis will allow to qualify analytically stationary configurations. We would like in the following to provide a more general analysis. 
Indeed, two questions remain at this step: given any pair of utility functions $(u_R,u_G)$, does a potential function exist and can we compute it? 
Reciprocally, given a potential function, can we find a pair of utility functions $(u_R,u_G)$ that can be translated into this specific potential function?

\subsection{Main result}
\indent Let us begin with some definitions.\\

\indent \underline{Definition}\\
\indent Let $\mathbb{U}$ be the set of pairs of utility functions $(u_R,u_G)$ that verify for all $(R,G)\in E_H$ the condition: 
\begin{equation} 
u_R(R,G)-u_R(R,G+1)=u_G(R,G)-u_G(R+1,G) \label{cond}
\end{equation}

\indent The condition \ref{cond} only imposes that if a block contains $R+1$ red agents and $G+1$ green agents, the utility gain a red agent would achieve if a green agent left must be the same than the utility gain a green would achieve if a red agent left.
The results in the following apply to pairs of utility functions verifying this condition. As we show below, this condition is not strongly restrictive, which means that our approach can be applied to virtually all the usual utility functions.\\

\indent \underline{Definition}\\
\indent Let $\mathbb{F}$ be the set of aggregate functions of the form $\mathcal{F}(x)=\sum_{q\in\mathcal{Q}}F(R_q,G_q)$, where $F$ is an intermediate function defined on the set $E_{H+1}$ of all possible numbers of red and green agents that can be present in a block.\\

\indent The main result of this chapter consists in the following claim:\\
\begin{center}
\begin{flushleft} \hspace{0.5cm} \underline{Claim 1}\\\end{flushleft}
\fcolorbox{black}{gris5}{
\begin{minipage}{1\textwidth}
To each aggregate function $\mathcal{F}=\sum_{q\in\mathcal{Q}}F(R_q,G_q) \in \mathbb{F}$ corresponds at least\footnote{Since the definition of the potential function makes only intervene its variation, $\mathcal{F}$ can be defined up to an additive constant. Similarly, a utility function can also be defined up to a constant. All the formula in this insert are written with the convention $u(0,0)=F(0,0)=F(0,1)=F(1,0)=0$. For more details, see the proof in appendix $A.2$.} one pair ($u_R,u_G$) of utility functions of $\mathbb{U}$ of which values can be expressed as:
\begin{equation}
\left \{ \begin{array}{l}
u_R(R,G) = F(R+1,G)-F(R,G)\\
u_G(R,G) = F(R,G+1)-F(R,G) \label{uF} \\
\end{array} \right.
\end{equation}
Reciprocally, for each pair of utility function $(u_R,u_G)$ of $\mathbb{U}$, there exists one\footnote{Same remark as before. Since $\mathcal{F}$ can be defined up to a constant, formally there exist more than one corresponding potential function.} corresponding potential function $\mathcal{F}_{[u_R,u_G]} = \sum_{q\in\mathcal{Q}}F_{[u_R,u_G]}(R_q,G_q) \in \mathbb{F}$ which can be expressed through the functional:
\begin{eqnarray}
 F_{[u_R,u_G]}(R,G)&=& \sum_{r=0}^{R-1}u_R(r,0) + \sum_{g=0}^{G-1}u_G(R,g) \label{Fu1}\\
&=&\sum_{r=0}^{R-1}u_R(r,G) + \sum_{g=0}^{G-1}u_G(0,g) \label{Fu2}
\end{eqnarray} 
where we use the convention that a sum is equal to zero when the index of its last term is strictly inferior to that of its first term (which happens here in practice when either $R=0$ or $G=0$). 
\end{minipage} }
\end{center}

\indent The complete proof of claim $1$ is given in appendix \ref{proof1}. 

\subsection{Interpretation}
\label{interpr}
\indent Condition \ref{cond} can be given two interpretations. 
In its original form, it says that there is a symmetry in the externality generated by green agents on red agents and by red agents on green agents: 
starting from a given neighborhood composition, the variation in utility produced by the departure of an agent of the other type must be the same for the two categories. 
Condition \ref{cond} can also be written as follows:
\begin{equation} 
u_R(R,G)+u_G(R+1,G)=u_G(R,G)+u_R(R,G+1) \label{cond2}
\end{equation}
which means that starting from any initial composition of a block, the sum of utilities of a red agent and a green agent entering successively in this block is the same whatever the order in which they enter. 
Both interpretations show that the value of function $F$ in a given neighborhood $q$ does not depend on the particular path of events that lead to the composition of this neighborhood.  
The same can be said of the sum of the $F$ intermediate functions, that is function $\mathcal{F}$. 

\indent In other words, it is always possible to define the intermediate function $F$ as corresponding to the variation of utility of a moving agent.
However, only a pair of utility functions verifying condition \ref{cond} allows this function to be path-independent and therefore uniquely defined for any given configuration.
As Eq \ref{Fu1} shows, this intermediate component of the potential function corresponds to the sum of utilities of the agents arriving in succession in the block. 
This sum is calculated starting from an empty block, agents being introduced one by one, first the red ones and then the green ones. 
As Eq \ref{Fu2} shows, the same sum is obtained if green agents are introduced first and red agents after. 
More generally, thanks to the utility functions verifying condition \ref{cond}, this sum is independent of the precise order in which the agents are introduced in each block. 

\indent $\mathcal{F}$ can therefore be interpreted as the sum of the incentives the agents had to move in the neighborhood where they are located. 
Indeed, if $x(t)$ denotes the state of the city at the iteration $t$, then the potential can be rewritten as 
\begin{equation}
\mathcal{F}(x(t))-\mathcal{F}(x(0)) = \sum_{t'=1}^{t}\Delta_{x(t'-1)x(t')} u
\end{equation}
where $\Delta_{x(t'-1)x(t')} u=0$ by definition if no move happens at iteration $t'$ and where we can take $\mathcal{F}(x(0))=0$ since the potential is defined up to a constant.
It could also be viewed as the minimum utility level each agent would require to accept quitting its neighborhood. 
As such, it represents, in the case $T\rightarrow 0$, the stability of the configuration $x$: the higher the potential function, the smaller the incentives for agents to move. 

\indent In summary, the main property of the potential function is that it reflects both the micro and macro scale. 
On the one hand, $\mathcal{F}$ is an aggregate function which only depends on the number $R_q$ and $G_q$ of red and green agents in each block. 
On the other hand, $\mathcal{F}$ also keeps tracks of the individual level since it corresponds to a sum of individual moves. When the stationary states are reached in the case $T\rightarrow 0$, $\mathcal{F}$ is optimized, which means that no agent can strictly improve her utility by moving.

\indent It is finally important to note the difference between the potential $\mathcal{F}$ and the collective utility $U$, which is the sum of the individual utilities experienced by the agents in a configuration $x$. Whereas $\mathcal{F}$ represents the sum of the agents' utilities at the time when they have moved into their current location starting with a totally empty city (or are considered to have done so), $U$ represents the sum of the agents' utilities once they are all settled. Hence, while stationary configurations maximize $\mathcal{F}$ they do not necessarily (and the following examples show that they generally not) maximize the collective utility.


\section{Applications}
\subsection{Reexamination of an historical example: Schelling utility function and the Duncan index}
\indent Suppose that the agents compute their utility with Schelling's utility function (which is equal to $1$ if their fraction of similar neighbors is superior or equal to $0.5$, and equal to $0$ otherwise). This utility function can be expressed in terms of the number of red and green neighbors as follows:
\begin{eqnarray} 
u_R(R,G) &=& \Theta(R-G) = \frac{1}{2}(1+|R+1-G|-|R-G|)\nonumber\\   
u_G(R,G) &=& \Theta(G-R) = \frac{1}{2}(1+|R-1-G|-|R-G|)\label{usch}
\end{eqnarray}
where $\Theta$ is the Heaviside function defined by: $\Theta(x)=0 \,\,\mbox{if } x < 0$ and $\Theta(x)=1 \,\,\mbox{if } x \geq 0$.\\
\indent It is easy to figure out that this particular pair of utility functions respects the condition \ref{cond}, and is therefore in the set $\mathbb{U}$. It is possible to compute the potential function rather directly thanks to its interpretation (as the sum of the utility of the agents being introduced one by one in the city, this sum being independent of the precise order of the introduction of the agents) we presented in previous section. To compute the potential of a given configuration $x\equiv \{R_q,G_q\}$, let us consider that we introduce in each block first the agents in majority (ie the red ones if $R_q > G_q$, the green ones if $G_q> R_q$, either the red or the green ones if $R_q=G_q$) and second the agents in minority. Each of the first agents has a utility of $1$ as he settles in the city (since his group is in majority in his block when he settles) while each of the other agents has a zero utility when he settles (since his group is in minority when he settles).\footnote{Notice that in this particular example, we do not use the convention $u(0,0)=0$. See appendix \ref{appli1} for more details.} Hence it is straightforward to write the potential as

\begin{eqnarray*}
\mathcal{F}(x) &=& const + \sum_{q\in\mathcal{Q}} \max(R_q,G_q)\\
&=& const +  \sum_{q\in\mathcal{Q}}\frac{1}{2}\Big( R_q + G_q + |R_q-G_q|\Big) \\
&=& const' +  \frac{1}{2}\sum_{q\in\mathcal{Q}} |R_q-G_q|
\end{eqnarray*}

One can verify that the same expression can be found using the relation \ref{Fu1} (see appendix \ref{appli1}), the computation being in this case more formal than what we present here.

\indent This potential function has an expression well known to scientists working on segregation. In the case where the total number of red and green agents in the city are equal ($N_R=N_G=N$), this potential function is indeed linear to the Duncan \& Duncan dissimilarity index defined by $D(x)=\frac{1}{2}\sum_q|R_q/N_R-G_q/N_G|=\frac{1}{2N}\sum_q|R_q-G_q|$, one of the first index of segregation that was proposed by \citep{duncan55}, which is equal to $0$ when the composition of each block reflects exactly the composition at the city scale and $1$ for completely segregated states.  To the best of our knowledge, an analytical connection between the two ``historical'' works of Schelling and Duncan \& Duncan on segregative phenomena such as the one provided by our model has never been found before.

\subsection{Potential function and collective utility}
\indent Generalizing \citep{zhang2004dmr}'s choice of utility functions, suppose that $u_R$ and $u_G$ are expressed as 
\begin{eqnarray} 
u_R(R,G) &=& aR + bG \nonumber\\ 
u_G(R,G) &=& bR + dG \label{uabd}
\end{eqnarray}
where $a$, $b$, $d$ are constant parameters\footnote{\citep{zhang2004dmr} took $b=d=-1$ and $a\geq -1$, the utility also including a fixed income term which makes it positive.}. One can easily verify that this particular pair of utility functions verifies the condition \ref{cond} and one can compute the corresponding potential function: 
\begin{eqnarray}
\mathcal{F}(x) &=& \frac{1}{2}\sum_q (aR_q(R_q-1) + dG_q(G_q-1) + 2bR_qG_q)\nonumber\\ 
&=& \frac{1}{2}\sum_q (R_q[a(R_q-1) + bG_q] + G_q[d(G_q-1) + bR_q]) \nonumber\\
&=& \frac{1}{2}\sum_q (R_q u_R(R_q-1,G_q) + G_qu_G(R_q,G_q-1)\nonumber\\
&=& \frac{1}{2}U(x)\\ \nonumber
\end{eqnarray} 
\indent In this particular case, the potential function is thus proportional to the collective utility. Reciprocally, one can verify (see proof in appendix \ref{proof_FU}) that if we want the potential function to be proportional to the collective utility, then the constant of proportionality is necessarily $1/2$ and the pair of the agents' utility functions necessarily takes the form \ref{uabd} (up to a constant).\\
\begin{figure}[h!]
\begin{center}
\includegraphics[width=14cm]{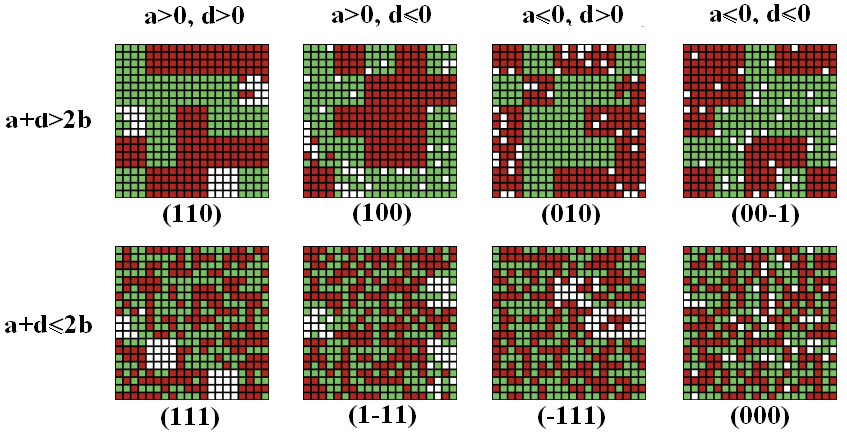}
\end{center}
\caption{\footnotesize{\textbf{Typical stationary configurations obtained by simulations for different values of ($adb$). Top panel:} for $2b-(a+d)<0$, the system evolves towards segregated configurations where red and green agents tends to live in different blocks. \textbf{Bottom panel:} for $2b-(a+d)>0$, the system evolves towards mixed configurations where the number of red-green pairs of neighbors is maximized. \textbf{From left to right:} the sign of $a$ and $d$ controls the tendency of red and green agent to prefer to live in dense or uncrowded areas. The demographic parameters are $(N=20,v=10\%,n_R=0.5)$. Neighborhood size is fixed to $H+1=9$ and the level of noise is $T=0.1$}}
\label{rhorhorho}
\end{figure}
\indent With this choice of utility functions, claim 1 and lemma 2 ensure that for low values of $T$, the stationary configuration will maximize the collective utility, but what about the segregation level? One can guess that for $a>b$ and $d>b$, the agents have strong preferences for all-similar neighborhoods, leading to highly segregated patterns, but we would like to be more specific. In order to achieve this goal, let us introduce 
\begin{eqnarray*}
\rho_{RG} &= \sum_q R_q G_q \hspace{2.5cm} &\mbox{the number of red-green pairs of neighbors,}\\ 
\rho_{RV} &= \sum_q R_q(H+1-R_q-G_q) &\mbox{the number of red-vacant pairs of neighbors and} \\ 
\rho_{GV} &= \sum_q G_q(H+1-R_q-G_q) &\mbox{the number of green-vacant pairs of neighbors.}
\end{eqnarray*}
Starting from these expressions, one can rewrite the potential function corresponding to \ref{uabd} as:
\begin{eqnarray}
\mathcal{F}(x) = \Big(b-\frac{a+d}{2}\Big)\rho_{RG}(x)-\frac{a}{2}\rho_{RV}(x)-\frac{d}{2}\rho_{GV}(x)
\end{eqnarray}

\indent This last form allows a convenient interpretation of the potential function. The term proportional to $\rho_{RG}$ gives a measure of the relative contact between the two groups, hence the sign of the prefactor $b-(a+d)/2$ indicates whether mixed states (when positive) or segregated states (when negative) are obtained at the global level. The terms proportional to $\rho_{RV}$ and $\rho_{GV}$ give a measure of the likelihood to find a vacant location around red and green agents. The prefactors $a/2$ and $d/2$ could hence be seen as a measure of what locations an agent can afford in term of income (supposing that the price increases with the local density). All these insights gained from the study of the potential function can be checked by means of simulations (Fig \ref{rhorhorho}).

\subsection{Review of some utility functions}
\label{pancs-functions}
\indent Another useful way to describe the set $\mathbb{U}$ is to remark that it is composed of the pairs of utility functions $(u_R,u_G)$ which are written
\begin{eqnarray}   
u_R(R,G) &=& \xi_R(R) + \sum_{g=0}^{G-1} \xi(R,g) \label{cond2a}\\
u_G(R,G) &=& \xi_G(G) + \sum_{r=0}^{R-1} \xi(r,G) \label{cond2b}
\end{eqnarray} 
where $\xi_R$ and $\xi_G$ are arbitrary functions of $\{0,1,..,H\}\rightarrow\mathbb{R}$ and $\xi$ an arbitrary function of 
$E_H \rightarrow\mathbb{R}$. With these notations, according to Eq. \ref{Fu1}, the potential function can be written as:\footnote{Notice that on this last form, it is pretty obvious that the potential does not depend on the order of arrival of the agents.}
\begin{equation} 
\mathcal{F}(x)=\sum_q\Big(\sum_{r=0}^{R_q-1} \xi_R(r) + \sum_{g=0}^{G_q-1} \xi_G(g) + \sum_{r=0}^{R_q-1} \sum_{g=0}^{G_q-1} \xi(r,g)\Big)
\label{Fxi}
\end{equation} 
\indent In the limit of a very low vacancy rate, there is no vacant cells in most of the blocks, \emph{ie} in these blocks the relation $R_q+G_q=H+1$ holds. Hence, one only needs one parameter among $(R_q,G_q,V_q)$ to define a utility function. Supposing that the utility of an agent only depends on his number of similar neighbors is then sufficient to describe all possible cases. This can simply be done by taking $\xi\equiv 0$ in Eq. \ref{cond2a} and Eq. \ref{cond2b}, while keeping the functions $\xi_R$ and $\xi_G$ independent and free. In other words, it is clear that in the limit of no vacant cells, each agent arriving in a neighborhood receives a utility that is fully determined by the number of like-neighbors. Therefore, the order in which the agents settle in the neighborhood does not matter and the condition for having a potential function holds. The set $\mathbb{U}$ hence describes all possible pairs of utility functions in the limit $v\rightarrow 0$.\\
\indent In the following, we place ourselves in the limit $v\rightarrow 0$. We take $\xi\equiv 0$ and review three pairs of utility function 
$(\xi_R,\xi_G)$ that have been studied elsewhere \citep{pancs07}. We will see that the potential function provides microscopic criteria for global segregation or for  integration.\\
\begin{figure}[h!]
\begin{center}
\includegraphics[width=14cm]{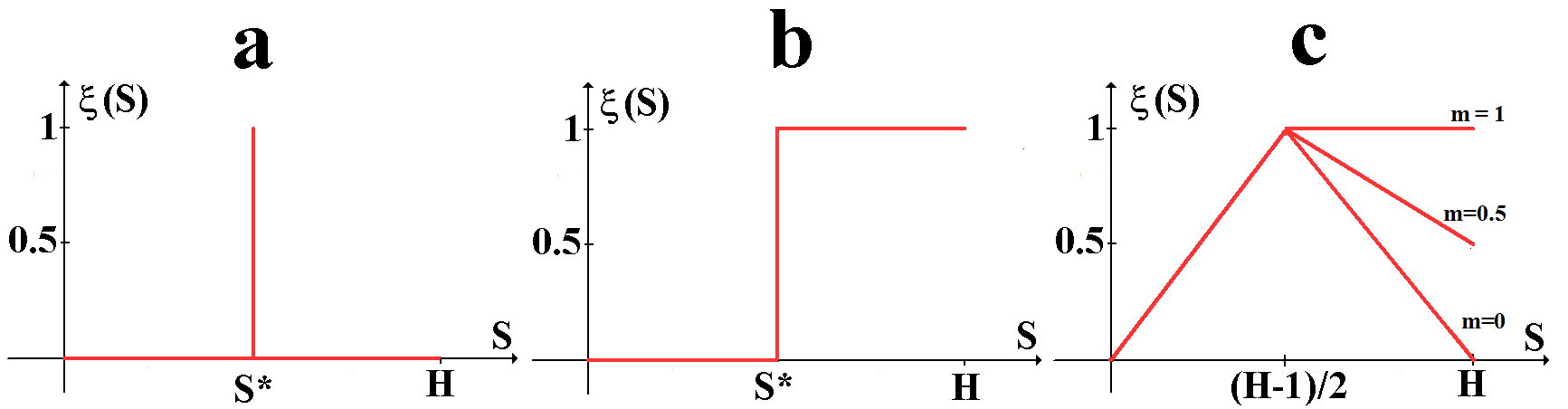}
\end{center}
\caption{\footnotesize{\textbf{Utility functions studied in this section.} 
\textbf{a.} Single-peaked utility function;
\textbf{b.} Schelling's utility function;  
\textbf{c.} the ``asymmetrically peaked'' functions.}}
\label{pancs-ufunctions}
\end{figure}

\subsubsection*{Single-peaked utility functions}
First, we consider the single-peaked utility function presented on Fig \ref{pancs-ufunctions}a. Here agents are driven exclusively by their obsession to a fixed number of similar neighbors $S^*$. The utility functions are given by
$$\left\{ \begin{array}{lll} 
\xi_R(R) &= 1 & \mbox{if $ R = R^* $ } \\ 
\xi_R(R) &= 0 & \mbox{otherwise} \\
\end{array} \right.
\left\{ \begin{array}{lll} 
\xi_G(G) &= 1 & \mbox{if $ G = G^*$} \\ 
\xi_G(G) &= 0 & \mbox{otherwise}  \\
\end{array} \right.$$
and a corresponding potential function is
\begin{equation}
\mathcal{F}(x)=\sum_q\Big(\Theta(R_q-R^*-1)+\Theta(G_q-G^*-1)\Big)
\end{equation}

We can then differentiate between the possible outcomes generated by a pair of single-peaked utility functions. Suppose for example that the two populations are equally present in the city ($N_R=N_G$), and that both populations have symmetric preferences ($R^*=G^*$). Remind that we placed ourselves in the limit $v \rightarrow 0$ and hence that the relation $R_q+G_q=H+1$ holds for almost all blocks $q$. For all intents and purposes, the potential function can thus be written  $\mathcal{F}(x)\simeq\sum_q\Big(\Theta(R_q-R^*-1)+\Theta(H-R^*-R_q)\Big)$. Hence,
\begin{itemize}
\item if $R^*\leq(H-1)/2$, it is possible to have both $R_q\geq R^*+1$ and $R_q\leq H-R^*$ in each block, a situation which clearly maximizes the potential. All configurations $x$ such that these inequalities are satisfied in every block are equiprobable. The probability for a block to contain $R_q$ red agents and $G_q$ green agents being proportional to the number $ (H+1)!/R_q!G_q!$ of ways of placing them, the distribution of the $R_q$ in the stationary states is a bell-like shaped curve restricted to $R^*+1\leq R_q \leq H-R^*$ and maximal for $R_q=(H+1)/2$. 
The red agents living in a block containing $R^*+1$ red agents and the green agents living in a block containing $G^*+1=H-R^*$ green agents have a utility of one, all the other agents have a zero utility. 
Hence both the mixity and the collective utility increase with $R^*$.
In the limit $R^*=G^*=(H-1)/2$, the perfectly mixed configurations are the only ones that maximize $\mathcal{F}$ and in those configurations the collective utility is also maximized.

\item if $R^* > (H-1)/2$, all configurations for which either $R_q\geq R^*+1$ or $R_q\leq H-R^*$ (these inequalities not being compatible anymore) in each block $q$ of the city maximize the potential and are equiprobable. In the stationary states, the distribution of the $R_q$ is always a bell-like shaped curve but restricted this time to $R_q \geq R^*+1$ and $R_q \leq H-R^*$ and maximal for $R_q=R^*+1$ and $R_q=H-R^*$. 
For $R^*=(H+1)/2$, each block always contain a majority group whose agents have a utility that depends on their precise number while the utility of the minority group is always zero. Hence the collective utility has dropped down compared to the $R^*=(H-1)/2$ case. 
As $R^*$ increases, the mixity now decreases while in the same time the collective utility increases.
In the limit $R^*=G^*=H$, the completely segregated configurations are the only ones that maximize $\mathcal{F}$ and in those configurations the collective utility is also maximized.        
\end{itemize}

The same kind of analysis allows to deduce the characteristics of the stationary states in the $R^* \neq G^*$ cases. What is interesting here is to notice that while maximal collective utility occurs in two cases ( $R^*=(H-1)/2$ with perfectly mixed configurations and $R^*=H$ with completely segregated configurations), only one of these cases is stable with respect to small changes in the agents' preferences. Starting with the $R^*=(H-1)/2$ case, small fluctuations in the agents preferences (them becoming slightly less tolerant) can induce a sharp transition in $U^*$ (from $1$ to $0.5$), while the stationary configurations remain almost unchanged. Imagining the preference of the unsatisfied agents can evolve, the simplest way for them to increase their utility is by becoming less tolerant inducing a loop of increase of segregation/decrease in tolerance/increase in utility. On the other hand, starting from the $R^*=H$ case, small fluctuations in the agents preferences induce only a small change in $U^*$ and the simplest way for the unsatisfied agents to regain their utility is to come back to $R^*=H$.

\subsubsection*{Stair-like functions}
Second, we consider a stair-like utility function, defined here through the number (and not the fraction\footnote{A pair of stair-like utility functions defined through the fraction of similar neighbors does not generally verify condition \ref{cond}, while it does when they are defined through the number of similar neighbors. }) of similar neighbors, presented on Fig \ref{pancs-ufunctions}b. The utility functions are given by
$$\left\{ \begin{array}{lll} 
\xi_R(R) &= 0 & \mbox{if $ R < R^* $ } \\ 
\xi_R(R) &= 1 & \mbox{if $R \geq R^* $ }  \\
\end{array} \right.
\left\{ \begin{array}{lll} 
\xi_G(G) &= 0 & \mbox{if $ G < G^* $ } \\ 
\xi_G(G) &= 1 & \mbox{if $G \geq G^* $ }  \\
\end{array} \right.$$
and a corresponding potential function is 
\begin{equation}
\mathcal{F}(x)=\sum_q\Big((R_q-R^*)\Theta(R_q-R^*-1)+(G_q-G^*)\Theta(G_q-G^*-1)\Big)
\end{equation}

We can again differentiate between the possible outcomes generated by a pair of stair-like utility functions. Taking once again equal populations with symmetric preferences, the potential function can also be reduced in the $v\rightarrow 0$ limit case to the simpler form $\mathcal{F}(x)\simeq \sum_q\Big((R_q-R^*)\Theta(R_q-R^*-1)+(H+1-R^*-R_q)\Theta(H-R^*-R_q)\Big)$.\\

For all values of $R^*$, the configurations which maximize this potential function are those for which either $R_q=0$ or $R_q=H+1$ in each block $q$ of the city. The system hence ends up in completely segregated configurations. Compared to the single-peaked utility functions case, the asymmetry towards like-neighbors in the utility function is favorable to segregated states and the normalized collective utility is always maximal in the stationary states.


\subsubsection*{Asymmetrically peaked functions}
Finally, we consider an asymmetrically peaked utility function presented on Fig \ref{pancs-ufunctions}c. For simplicity, we suppose that the number  $H$ of possible neighbors of an agent is odd. The utility functions can thus be written as ($S$ standing for either $R$ or $G$)
$$\left\{ \begin{array}{lll} 
\xi_{S}(S) &= 2S/H & \mbox{if $S\leq (H-1)/2 $} \\ 
\xi_{S}(S) &=m + 2(1-m)(H-S)/H &\mbox{if $S>(H-1)/2$ } \\
\end{array} \right.$$

and a corresponding potential function can be written as $\mathcal{F}(x) = \sum_q F(R_q,G_q) = \sum_q \Big(\tilde{F}(R_q) + \tilde{F}(G_q)\Big)$,
where 
\begin{equation}
\tilde{F}(S) = \big(S-\frac{H-1}{2}\big)\big( S-\frac{H+1}{2}\big) \Big[\frac{1}{H-1}\Theta \big(\frac{H+1}{2} - S\big) -\frac{1-m}{H+1}\Theta \big( S-\frac{H+3}{2}\big)\Big]
\end{equation}

This expression of the potential put forward the crucial role of the asymmetric parameter $m$. Indeed, one can compare different repartitions of $H+1$ red and $H+1$ green agents in two blocks of the city thanks to the potential function $\mathcal{F}$. Taking $x=R_q - G_q$ the difference in the number of agents of the two categories in the two neighborhoods, the difference in the potential function compared \emph{ceteris paribus} to the even distribution of agents in these two neighborhoods can be written:
\begin{eqnarray}
\Delta \mathcal{F} = F\big(\frac{H+1}{2}+x,\frac{H+1}{2}-x\big)+F\big(\frac{H+1}{2}-x,\frac{H+1}{2}+x\big) - 2F\big(\frac{H+1}{2},\frac{H+1}{2}\big) =& \nonumber\\
\frac{2}{H^2-1}\big(m(x^2+x)(H-1)+2x^2-2Hx \big) \hspace{3cm} & \label{rr}
\end{eqnarray}

\indent As could be expected, this expression increases with $m$, which means that a given segregated configuration is more probable and stable as the asymmetry toward like-agents is increased. It also increases with $x$, which means that for a given $m$ a highly segregated block is more probable than a scarcely segregated one. 
More importantly, a highly segregated block will be more probable than a perfectly mixed one if and only if the relation \ref{rr} is positive for $x=(H+1)/2$, which can be rewritten as 
\begin{equation}
m>\frac{2}{H+3}
\end{equation}
\indent Our analysis hence provides a microscopic criteria allowing to predict a global outcome. For $m>2/(H+3)$, complete segregated configurations will be obtained at the expense of the collective utility and for $m<2/(H+3)$, perfectly mixed configuration will be obtained. These results hold of course in the limit of low noise ($T\rightarrow 0$). It is remarkable that our analytical model predicts here a critical value of the asymmetric parameter $m$ that is qualitatively comparable to the critical value we obtained in the more classical model of chapter $1$ (we obtained a critical value $m\simeq 0.35$ with continuous neighborhood, $H=8$, a vacancy rate $v=10\%$ and a noise level $T=0.1$, see Fig. \ref{dyn-uap}). Our analytical results however depend on the precise definition of the model. It can be shown that taking the peak of the utility function to $H/2$ instead of $(H-1)/2$ leads to a negative critical value of $m$. However, in all those different models (analytical ones and simulated ones) the critical value of $m$ converges towards $0$ as the size of the neighborhood $H$ is increased.\\


\section{Extensions}
To this point, we have focused our attention on a formal version of a classical Schelling-type segregation model. But our approach provides an analytical framework that allows to investigate many extensions and to consider a broader range of issues than other classical approaches. Some preliminary suggestion yet to be fully developed are presented in this section.

\subsection{Segregation by ethnic origin, income, and preferences for public amenities}
\indent Until now, we have always implicitly supposed that the sole characteristic that the agents use to evaluate a location is the composition of its neighborhood. The red and green labeling of our two groups thus corresponds to two groups of different ethnic origin or two groups with different average income level. 
Other determinants of residential location choice however exist and are not correlated to the ethnic origin or the social economic status. Indeed, \cite{tiebout56}'s analysis of the importance of the location of local public goods is perhaps the main competitor of \cite{schelling71} in terms of its influence on later work on neighborhood choice.\\

It is very easy to write versions of our model which take into account the agents' public good preferences while keeping the existence and properties of a potential function. Noting for example $A$ the set of all the public amenities (city center, supermarkets, schools, ...) and $d_{i,a}$ the distance between an agent $i$ and an amenity $a\in A$, the utility of an agent $i$ could be rewritten in a general fashion as
\begin{equation}
u_i(R,G)+\tilde{u}_i(\{d_{i,a}\}_{a\in A})
\end{equation}
and one could easily derive the more general form of the potential function
\begin{equation}
\mathcal{F}(x)+\sum_i \tilde{u}_i(\{d_{i,a}\}_{a\in A})
\end{equation}

This generalized approach could provide a means to correct one of the bias of our analytical model, namely the lack of heterogeneity between 
the agents.
However, the extraction of the properties of the stationary states from this condensate global function would become quite challenging, as the dimension of the state variable of the system increases with the number of added amenities.

\subsection{Taxation to sustain collective welfare}
The basic concept at the center of a Schelling model is that of an agent deciding where to move according solely to the benefit $\Delta u$ she would achieve if she was to move. Her move affecting her past and new neighbors, an implicit consequence is that she could generate externalities that amount to $\Delta U -\Delta u$ while moving. 

Suppose now the existence of a benevolent planner who rewards positive externalities and taxes negative externalities. A way to model the action of that hypothetic benevolent ruler is to write the probability that a move happens as:
\begin{equation}
Pr\{move\} = \frac{1}{1+e^{\,-\big(\Delta u + \alpha (\Delta U -\Delta u)\big) / T}}
\end{equation} 
where $0 \leq \alpha \leq 1$ is a parameter controlling the tax level. The limit case $\alpha=0$ corresponds to a standard Schelling model and the limit case $\alpha=1$ corresponding to a case where only the interest of the collectivity as a whole is taken into account.

Following the path of the proofs developed in section $2.2$, one can infer the stationary distribution in the bounded neighborhood framework:
\begin{equation}
\Pi(x)=\frac{e^{\big((1-\alpha)\mathcal{F}(x)+\alpha U(x)\big)/T}}{\sum_z e^{\big((1-\alpha)\mathcal{F}(z)+\alpha U(z)\big)/T}}
\end{equation}

The potential function can thus in this context be generalized to $(1-\alpha)\mathcal{F}(x)+\alpha U(x)$. We already noted that the configurations maximizing $F$ are not in general maximizing $U$ and could even in certain cases (asymmetrically peaked utility function) be very unfavorable to $U$. In this context, the question of interest is to determine what level of tax $\alpha$ is necessary or sufficient to break undesired stationary configurations obtained in the classical Schelling model. Such questions are addressed analytically in another paper \citep{econophys-unpublished} and by means of simulations in the next chapter.


\chapter{Effects of coordination}
\section{Introduction}

\indent One of the most important features of Schelling-type models is that their dynamics is governed by the agents' individual preferences. In this context, a presumed condition for integration to occur is that the agents have a preference for a mixed environment. But as we saw in several examples in the previous chapters, that condition is often not a sufficient one, mainly because mixed configurations are unstable with respect to fluctuations, whereas segregated configurations are very stable.\\
\indent The stability of the city configurations, and moreover the gaps between the agents' micro-motives and the emergent macro-behavior, are linked to the externalities generated by the moving agents. This is particularly the case when the agents preference are given by the asymmetrically peaked utility functions, the system ending in highly segregated configurations whereas the agent's main preference is for mixed neighborhoods. A way to solve the integration issues in this case could thus be to focus on the welfare issues, \emph{ie} to find some mechanism to reduce the externalities generated by the agents' movements.\\  
\indent A classical idea in economics to avoid the impact of the externalities generated by individual selfish moves is to impose a taxation equal to the externality his move generates. Such a policy implemented by a central authority is known to lead to first-best equilibria, where the collective utility reaches its highest possible value. We verify here in section $3.3.$ that this is indeed the case in the context of Schelling's model. We also investigate the impact of different levels of taxes and we show that a tax equal to only one fifth of the generated externalities is sufficient in certain cases to reduce consequently the gap created by the agents' selfish behavior. To the extent of our knowledge, this work has never been done in the context of Schelling segregation models.\\
\indent However, these tax policies are not easy to establish in practice since they require a very accurate knowledge of the city at the local scale by the central authority. Such information is rarely perfectly and costlessly available. Hence, the policies implemented by central authorities correspond generally to mechanisms that can only reach second-best equilibria. Recently, some papers have proposed to add some `second-best' mechanisms in the original Schelling model to reinforce the integrated configurations. \cite{dokumaci07} proposes to tax the agents proportionally to the density of population in their neighborhood. The tax level depends on the ethnic origin of the agents, as might be the case under various form of affirmative action. \cite{barr} introduces additional social interactions into the Schelling model by coupling it with a Prisoner's Dilemma played with neighbors. In both case however, the effect of the added mechanism could be reformulated in terms of a redefinition of the agents' utility function.\\
\indent We propose in section $3.4$ to investigate the effect of the introduction of a local coordination by vote of co-proprietors. That new coordination mechanism has the advantage of remaining in the spirit of Schelling's model, adding only individual decisions based on the same utility as the moving agent, without any need of a central authority.\\

Before developing models with coordination or tax, we present in section $3.2.$ the standard Schelling model which we will use afterwards as reference. \\


\section{A standard model}
\subsection{Basic setup}
In all the simulations presented in this chapter, the demographic parameters of the city are fixed. The size of the city is set to $N^2=400$, a good compromise between the necessity to take a large value of $N$ to avoid small city effects\footnote{Such as those emphasized by \cite{singh2007}.} and the convenience to take a small value of $N$ to achieve short computation times. The number of agents of each group is fixed to $N_R=N_G=180$ and the vacancy rate is fixed to $v=10\%$. As usually assumed, we use continuous neighborhoods, that is, the neighbors of an agent are the agents living on the $H$ nearest cells surrounding him. Unless otherwise stated, the neighborhood size is fixed to $H=8$.\\
\indent In order to simplify our study, we will suppose that all the agents share the same ``asymmetrically peaked utility function'' $u_m(s)$, where $s$ is the fraction of one's similar neighbors and $0\leq m \leq 1$; $u_m$ is defined by:
\begin{eqnarray*}
u(s) &=& 2s \hspace{3.4cm}\mbox{for $s\leq 0.5$} \\
u(s) &=& m + 2(1-m)(1-s)\hspace{0.5cm}\mbox{ for $s>0.5$}
\end{eqnarray*}

\begin{figure}[h!]
\begin{center}
\includegraphics[width=7cm]{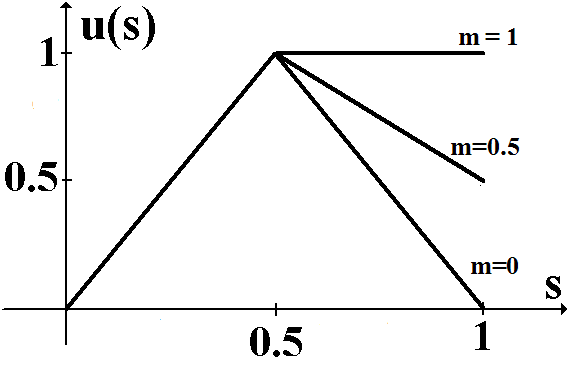}
\end{center}
\caption{\footnotesize{In our simulations, the agents all share the same utility function $u_m(s)$. We explore the change in behavior of the agents according to the value of the parameter $m$. Except for $m=1$, the agents always have a strict preference for perfectly mixed neighborhood. Except for $m=0$, their utility function presents an asymmetry: they prefer all-similar neighborhoods to all-dissimilar neighborhoods.}}
\label{uapuap}
\end{figure}

\indent By varying the value of the parameter $m$, we will then be able to explore the responses of our system to a whole family of utility functions. Our choice of working only with the asymmetrically peaked utility function is driven by the observation we made in chapter 1 (which is supported by the analytical results of chapter 2):  for $m$ roughly superior to $0.3$, the asymmetry in favor of the all-similar neighborhood in these utility functions leads to segregation patterns at the city scale at the cost of a low collective utility. This family of utility functions is thus the perfect candidate for testing whether the introduction of any type of coordination might allow to break segregative patterns and lead to more integrated patterns in which the collective utility would be higher.

\subsection{Dynamic rule}
\indent As explained before (section 1.2), in standard Schelling-type models agents move only to satisfy their own interest. We suppose that the dynamic follows a logit behavioral rule: at each iteration, an agent and a vacant cell are randomly chosen and the probability that this picked agent moves in that vacant cell is written as:  
\begin{equation}
Pr\{move;\, WC\} = \frac{1}{1+e^{\,-\Delta u / T}}
\end{equation}
where WC stands for ``Without Coordination''. As showed in section 1.4, the logit dynamic rule has the advantage to grasp more aspects of reality (taking into account idiosyncratic amenities) and to lead the system to stationary results which are independent of the initial configuration. This is why we prefer that kind of dynamics to a more standard ``best response'' dynamic rule.

\subsection{Simulations}
\indent We introduce the parameter $\tau$ as the average number of moves per agent. Considering the demographic parameters we use in our simulations, an increment of $1$ in $\tau$ corresponds to $(1-v)N^2=360$ performed moves. In the simulations presented below, we use $\tau$ as a chronological reference\footnote{A more obvious choice of chronological reference could be the number $t$ of simple iterations (i.e. the number of attempted moves). Neither choice accounts for the proportion of accepted moves (whose cumulated value is $\tau/t$), whose instantaneous value depends on the values of the various parameters and on the state of the system. Ideally, it may be interesting to follow this rate of ``moving iterations'' with the dynamic evolution of the city.}. 

\subsubsection{First example}
\begin{figure}[h!]
\begin{center}
\includegraphics[width=13cm]{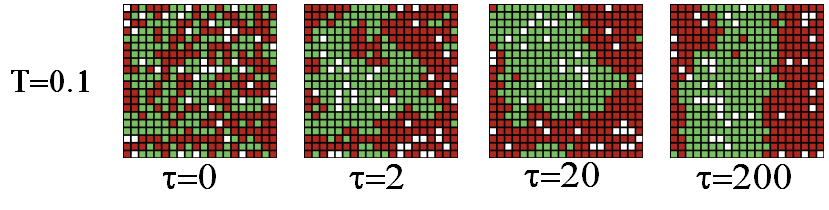}
\includegraphics[width=13cm]{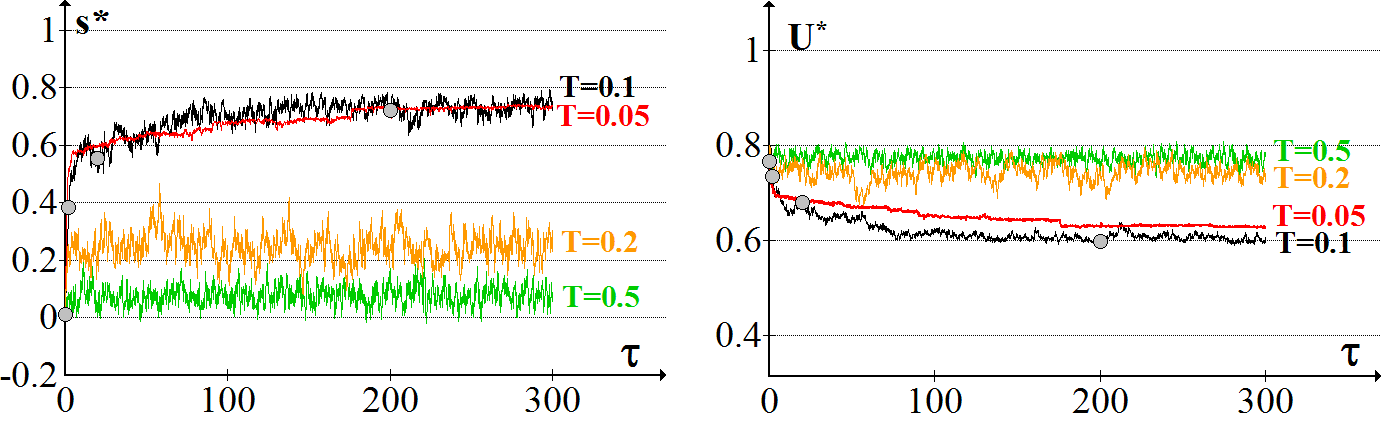}
\end{center}
\caption{\footnotesize{ Evolution towards a highly segregated configuration starting from a random configuration in the case of the WC rule. \textbf{Top panel.} Some snapshots of the evolution of the city for a noise level $T=0.1$; \textbf{Bottom panel.} Evolution with $\tau$ of the index of similarity and of the collective utility for different simulations with different noise levels. The grey dots on the $T=0.1$ curves correspond to the snapshots presented on the top panel. $m=0.5$.}}
\label{WC}
\end{figure}
\indent We present on fig \ref{WC} a typical evolution of the city in the case where the agents move without coordination, the level of noise being fixed to $T=0.1$ and the parameter $m$ to $0.5$. Starting from a random configuration, we observe the rapid formation of homogeneous areas which slowly melt into one another leading to the emergence of a highly segregated configuration where the city is divided into two uniform areas, each  inhabited by only one type of agent. 
The bottom panel of fig \ref{WC} shows that after a transition time, the city enters a stationary phase in which the segregation index $s^*$ and the (normalized) collective utility $U^*$ fluctuate with rather low amplitudes. Even though the agents' preferences go to mixed configurations and even though the agents are prompted to move in order to improve their own utility, their moves lead to a highly segregated configuration at the city level (the stationary value of $s^*$ is close to $0.8$) in which most of the agents are far from being fully satisfied (the stationary value of $U^*$ is close to $0.6$). This evolution is similar to those presented in Chapter 1.

\subsubsection{Influence of the parameters $m$ and $T$}
\begin{figure}[h!]
\begin{center}
\includegraphics[width=7cm]{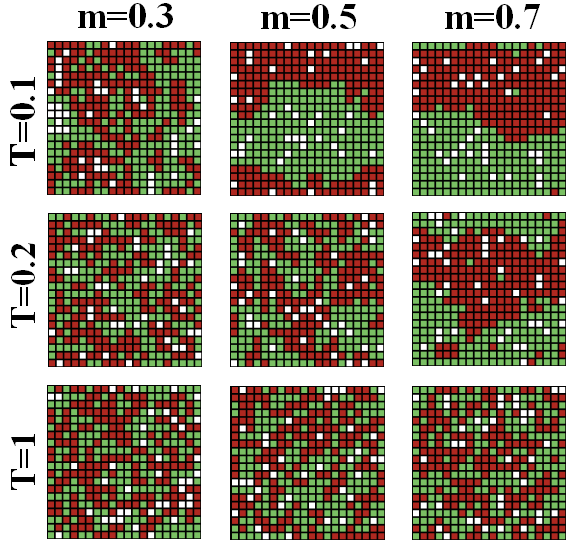}
\end{center}
\caption{\footnotesize{\textbf{Typical stationary configurations obtained with the WC dynamic rule for different values of $m$ and $T$.}}}
\label{WCstudy}
\end{figure}

\begin{figure}[h!]
\begin{center}
\includegraphics[width=6.5cm]{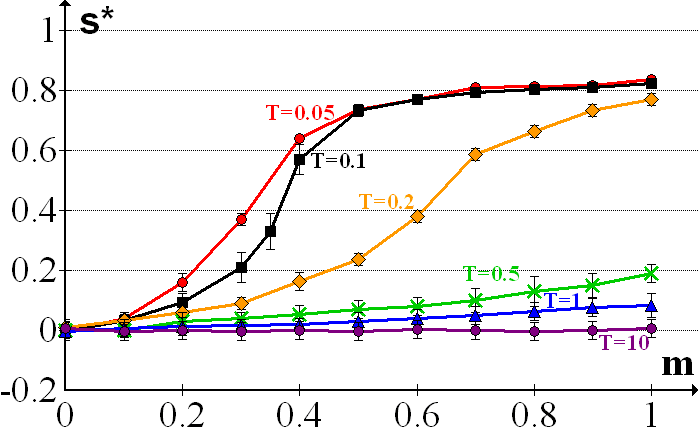}
\includegraphics[width=6.5cm]{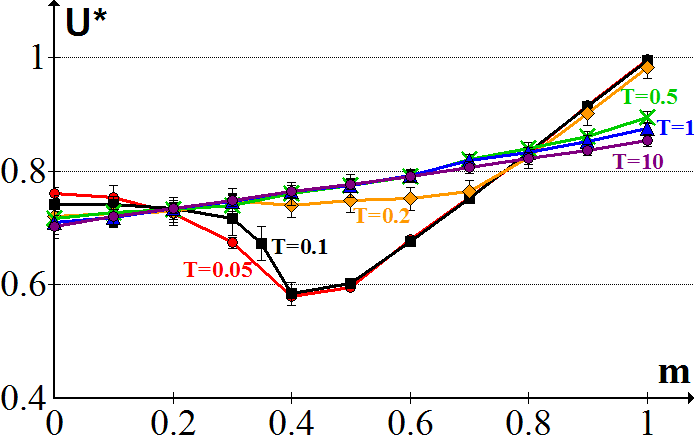}
\end{center}
\caption{\footnotesize{\textbf{Stationary mean values of the similarity index and of the collective utility} as a function of $m$, for different level of noise $T$. The error bars give the standard deviation of the fluctuations of $s^*$ and $U^*$ once the system has reached its stationary phase. The mean value and standard deviation are computed over $10$ periods of the stationary phases (when the fluctuation are important, we use larger temporal windows).}}
\label{WC2}
\end{figure}

\indent The influence of the parameters $m$ and $T$ can be observed on Figs. \ref{WC} to \ref{WC2}. For high values of $T$ ($T \geq 0.5$), the dynamic is essentially governed by the randomness introduced in the logit. In the limit $T \gg 1$, the agents are distributed uniformly in the corresponding stationary configurations, which induces a similarity index equal to zero for all values of $m$. The probability $p(s)$ for an agent to have a fraction $s$ of similar neighbors being independent of $m$ in the limit $T\gg 1$, the collective utility can thus be written (the sums being taken on all the discrete possible values of $s$):
\begin{equation}
U = \sum_{s} p(s)u(s) = \cdots = U_{m=0} + m \sum_{s \geq 0.5}(2s-1)p(s) 
\end{equation}  
The bilinear form of the asymmetrically peaked utility function induces the linear dependency of the collective utility with $m$ observed on fig \ref{WC2} for high values of $T$. Indeed, there are two parts in $U^*$ when $T$ is large and the distribution of agents almost random: because the distribution does not differ from the distribution when $m=0$, agents have the same ``baseline'' utility ; however, those who have more than half of same-type neighbors have a higher utility level than in the $m=0$ case, and the gap is linearly increasing in $m$. \\
\indent For low values of $T$ (roughly $T \leq 0.1$), the results are similar to the ones we presented in section \ref{dyn-issues} in chapter 1. The level of noise being low, the dynamic is governed mainly by its deterministic part, \emph{i.e}, the agents' preferences. One can refer to section \ref{dyn-issues} for a detailed interpretation of the dependency of the results with $m$. What is interesting to notice here is that for middle values of $m$ (between $m=0.2$ and $m=0.8$) the final outcome for low values of $T$ is always worse from a welfare point of view than it is in a random distribution of the agents. This observation clearly points out the deficiency with respect to social welfare of the location mechanism.\\ 
\indent The level of noise also has an effect on the fluctuations of $s^*$ and $U^*$ in the stationary phase as can be observed on fig \ref{WC}. While these fluctuations are rather low for $T=0.05$, their amplitude increases with $T$ before reaching a saturation value. Finally, while further studies would be necessary to characterize the influence of $T$ on the time needed to reach the stationary phase, one can infer from \ref{WC} that as $T$ decreases, this transition time increases (see lemma 1 in section \ref{abbr}).   

\indent The next two sections present two different ways of introducing coordination in the model in order to explore the robustness of the deficiency of the location mechanism with this lack of coordination. 


\section{A partial coordination by taxation}
\subsection{Basic setup}
The idea of introducing 'partial coordination' is to take into account a fraction of the externality generated by a moving agent on all the affected agents, \emph{ie} her past and potentially new neighbors. A mechanism of this kind can be interpreted as the intervention of a benevolent planner who taxes negative externalities and rewards positive externalities.\\

\subsection{Dynamic rule: tax on the externalities}
\indent In our reference case (without coordination), an agent decides to move according solely to the benefit $\Delta u$ he would achieve if he was to move. As a consequence from that move, she could generate externalities that amount to $\Delta U -\Delta u$. According to our premise that a benevolent planner would reward positive externalities and tax negative externalities, we propose to write the probability that a move happens by modifying the WC dynamic rule as follows:
\begin{equation}
Pr\{move; PC\} = \frac{1}{1+e^{\,-\big(\Delta u + \alpha (\Delta U -\Delta u)\big) / T}}
\end{equation} 
where PC stands for ``Partial Coordination''. $0 \leq \alpha \leq 1$ is a parameter controlling the tax level, the limit case $\alpha=0$ corresponding to the WC case and the limit case $\alpha=1$ corresponding to a 'Global Coordination' case where only the interest of the collectivity is taken into account.\\ 
\indent From an analytical point of view, these changes clearly do not affect the main property of the Markov chain theory: there exists one unique stationary distribution and hence the independence of the final configurations on the initial ones is still valid. For $\alpha=1$, the probability to move involves only the global function $U$. It is pretty easy to figure out - following the path developed in section \ref{def-prop} - that the stationary distribution can be written as :
\begin{equation}
\Pi_{PC,\alpha=1}(x)=\frac{e^{U(x)/T}}{\sum_{z \in X} e^{U(z)/T}}
\end{equation} 
Hence, according to lemma $2$, the configurations obtained in the limit $T\rightarrow 0$ are those which maximize the collective utility.\\
\indent In section $2.5$, we found an analytical expression of the stationary distribution for other values of $\alpha$ in the context of a bounded neighborhood. For continuous neighborhoods, our analytical approach is no longer valid. The reason can be stated quite simply : in the bounded neighborhood case, the information used to calculate the utility difference achieved by the moving agent (the initial and final densities) allows to calculate the difference of the global utility. This is because the agent's initial neighbors share the same neighborhood as him, and therefore a utility difference that can be calculated, the same being true for the final neighbors. Instead, in the continuous neighborhood, the global utility difference depends on the neighbors of the neighbors of the moving agent. Indeed, the utility difference felt by a neighbor of the moving agent depends on his own neighbors, most of which are not neighbors of the moving agent. Lacking the analytical approach, one still needs to turn to simulations in order to investigate the effects of the introduction of partial coordination.

\subsection{Simulations}
\begin{figure}[h!]
\begin{center}
\includegraphics[width=10cm]{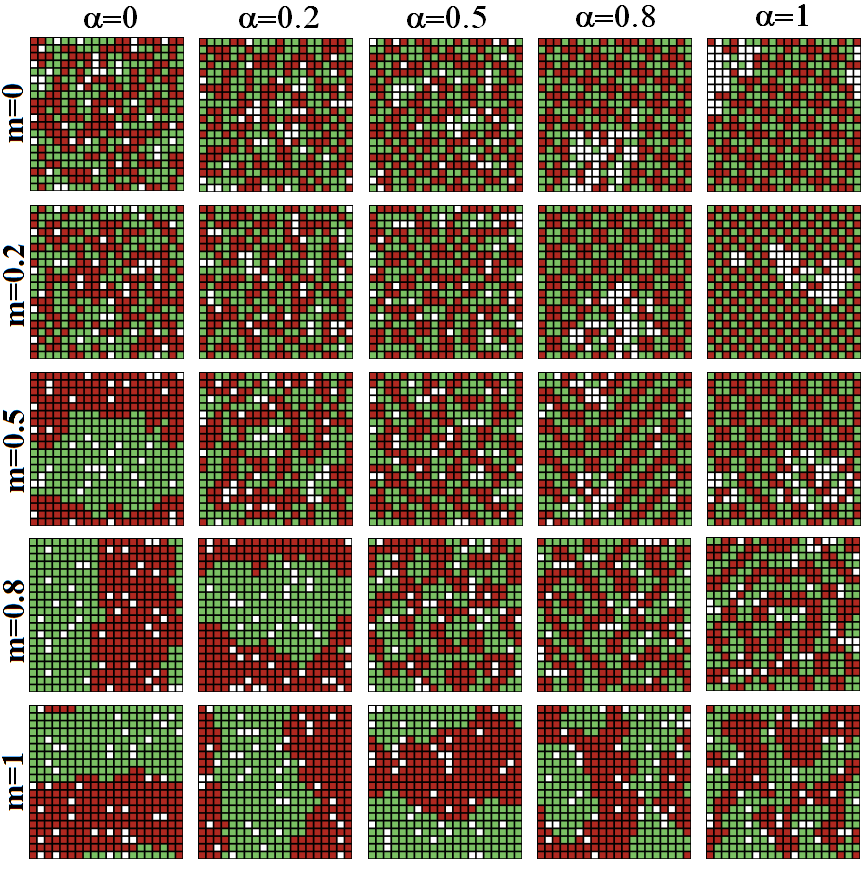}
\end{center}
\caption{\footnotesize{\textbf{Snapshots of typical stationary configurations} of the city for different values of $\alpha$ and $m$. The introduction of partial coordination can destabilize the highly segregated configurations and leads to configurations which present structured mixed patterns. $T=0.1$ and $H=8$.}}
\label{PCsnap}
\end{figure}

\begin{figure}[h!]
\begin{center}
\includegraphics[width=6.5cm]{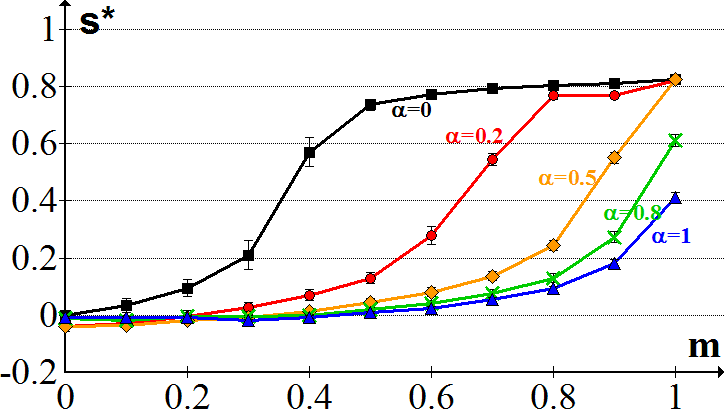}
\includegraphics[width=6.5cm]{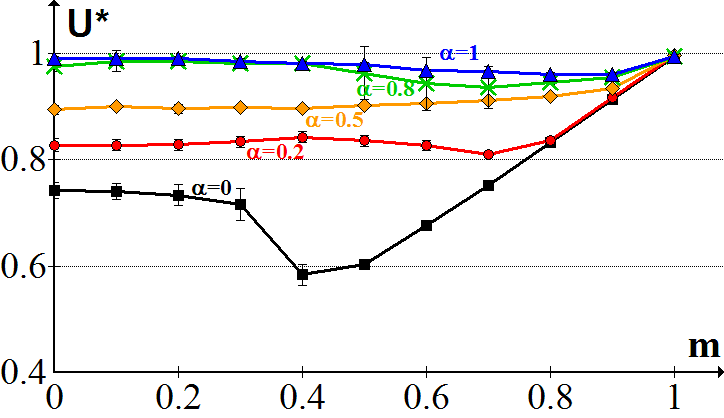}
\end{center}
\caption{\footnotesize{\textbf{Stationary mean values of the similarity index and of the collective utility} as a function of $m$, for different tax level $\alpha$, with the partial coordination dynamic rule. The error bars give the standard deviation of the fluctuations of $s^*$ and $U^*$ once the system has reached its stationary phase. The mean value and standard deviation are computed over $10$ periods of the stationary phases. $T=0.1$ and $H=8$.}}
\label{PC}
\end{figure}

\indent We present on Figs. \ref{PCsnap} and \ref{PC} snapshots of typical stationary configurations along with the corresponding values of $s^*$ and $U^*$  for different values of $\alpha$ and $m$. As previously stated, the case $\alpha=0$ corresponds exactly to the WC case. For $m \leq 0.3$, the preference for a mixed neighborhood of the agent prevails (locally mixed configurations are observed) and the segregation index is low. The dispersion in the distribution of neighborhoods' composition induces a level of mean utility around $0.75$, which is just a bit better than what is obtained with a random allocation of agents (see fig \ref{WC2}). For $m\geq 0.4$, the asymmetry in the agents' utility function induces a higher stability of highly segregated states to which the system converges. These segregated states are particularly harmful in terms of welfare for middle value of $m$ $0.4\leq m \leq 0.7$.\\

\indent Note first that adding coordination with $\alpha=1$ to the $m=0$ case shifts the random configuration to an ordered one: enhancing the utility level is possible only by achieving $s=0.5$ in every location.
Note also that even if a tax in this case does not affect the value of the similarity index, it allows to enhance welfare by clustering vacancies, diminishing the number of agents that do not have $s=0.5$ exactly.
For intermediate values of $m$, increasing $\alpha$ breaks segregation patterns.
Figure \ref{PC} shows that even a tax equal to one fifth of the generated externalities is enough to change the segregation level and to increase utility for $0.4\leq m \leq 0.7$.
Obviously, for high values of $m$, due to the form of the agents' utility function, a high utility level is obtained whatever the tax level. 
Still, the similarity index is lower when coordination is introduced: changing $\alpha$ from 0.5 to 0.8 and then to 1 decreases $s^*$.\\

Economic theory predicts that optimality (a collective utility of $1$) is obtained if a tax equal to the generated externalities is implemented. For $\alpha=1$, this is what would be obtained in the limit $T=0$. Here, the collective utility is slightly inferior to $1$ because of the finite value of the noise ($T=0.1$). Interestingly, a low level of taxation is able to significantly increase welfare. A tentative explanation is that the number of affected agents being in general greater than one, the variation in collective utility would on the average be greater than the variation of the moving agent ($|\Delta U|/|\Delta u|$ being in this hypothesis proportional to $H$). This argument is in fact incorrect because, while the moving agent potentially undergoes a big change of neighborhood ($|\Delta u| \sim 1$), the neighborhoods of the affected agents only slightly change ($|\Delta U|\sim H\cdot 1/H\sim 1$). A possible explanation of why a low tax is sufficient is that it is based on cumulative mechanisms similar to the ones that lead to segregation in the WC case. 

\section{A local coordination by voting}
\subsection{Basic setup}
We define the \emph{co-proprietors} of an agent as the agents living on the $h$ nearest cells surrounding him. Here, $h$ is a fixed integer which verifies $h \leq H$. \emph{Co-proprietors} represent in a stylized way next-door neighbors or the people living in the same residential building whereas the \emph{neighbors} represent the people living in the same street or in the same district. Examples of possible forms of neighborhoods and co-properties are shown below in Fig \ref{neighbor+copro}.\\
\begin{figure}[h!]
\begin{center}
\includegraphics[width=9cm]{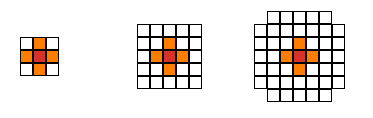}
\end{center}
\caption{\footnotesize{\textbf{Examples of neighborhood and co-property used in our simulations.} A red agent is located on the central cell. His co-property corresponds to the orange cells and his neighborhood to the orange and white cells. From the left to the right, $(H=8,h=4)$, $(H=24,h=4)$ and $(H=44,h=4)$.}}
\label{neighbor+copro}
\end{figure}
\indent We introduce local coordination by taking into account the potential change of utility of the co-proprietors of the vacant cell considered by the potential mover. In the following, we will denote by $\mathcal{C}$ this set of agents. It seems more logical to introduce local coordination through the potentially new co-proprietors (who have you take an admission exam) than through the current co-proprietors (that you can quit on your free will). For mathematical convenience, we will suppose that the probability that the move happens can be computed as the product of the probability that the potential mover would like to move and the probability that the agents of $\mathcal{C}$ accept him. The `local' nature of the implied coordination comes from the fact that only a fraction of the agents who might be affected by the potential move are consulted.\\
\indent We propose here one dynamic rule to counterbalance the wish of the potential mover by the opinion of his potentially new co-proprietors. Other mechanisms can of course be imagined.\\

\subsection{Dynamic rule: qualified vote of the co-proprietors}
\indent The simplest local coordination rule is that the potential mover needs the majority of the co-proprietors to endorse his moving in. Let $\Delta u_i$ be the variation of utility of the co-proprietor $i \in \mathcal{C}$ if the move was to take place. We write the probability that the co-proprietor $i$ votes `for' the move: 
\begin{equation}
Pr\{i,\mbox{`for'}\}=\frac{1}{1+e^{\,-\Delta u_i / T_2}}
\end{equation}
\indent The logit form of this acceptation probability can be justified in the same way as the logit for the moving agent (appendix \ref{Alogit}). The parameter $T_2>0$ can then be interpreted as the amplitude of a noise that represents in a stylized way the preferences of the co-proprietors over any characteristics of the potential mover other than and not correlated to the group he belongs to (marital status, number of children, profession, religion, friendship, etc...). Since a co-proprietor does not move, the noise affecting him is different from the noise affecting a moving agent : the arrival of a new neighbor does not change his position relatively to the city center, the supermarket, his children's school or any other amenities. Hence the level of noise $T_2$  is by nature very different from the level of noise $T$.  We moreover argue that since a co-proprietor deciding whether to accept or not a new neighbor is qualitatively subject to less characteristics other than the group membership (\emph{ie} to less noise) than a moving  agent, it is realistic to suppose $T_2 \leq T$.

\indent The move takes place with a probability: 
\begin{equation}
Pr\{move; LC\} = \frac{1}{1+e^{\,-\Delta u / T}}\,Y\Big(\Big\{\frac{1}{1+e^{\,-\Delta u_i /T_2}}\Big\}_{i \in \mathcal{C}}\Big)
\end{equation}
where $Y=1$, $1/2$ or $0$ if respectively more than half, exactly half or less than half of the co-proprietors vote `for' the move, and where LC stands for ``Local Coordination''.\\

\indent From an analytical point of view, the introduction of the vote of the co-proprietors does not change the main property of our system: it can always be described as a Markov chain, and since $T_2>0$, every move still has a non-zero probability to happen. This ensures the existence of one unique stationary distribution and hence the independence of the stationary states with regards to the initial starting state.\\

\subsection{Simulations}
In the following, we limit our investigations by fixing the noise level $T$ to $0.1$.

\subsubsection{First example}
\begin{figure}[h!]
\begin{center}
\includegraphics[width=13cm]{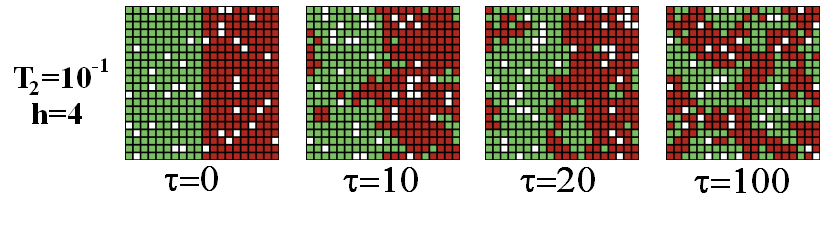}
\includegraphics[width=13cm]{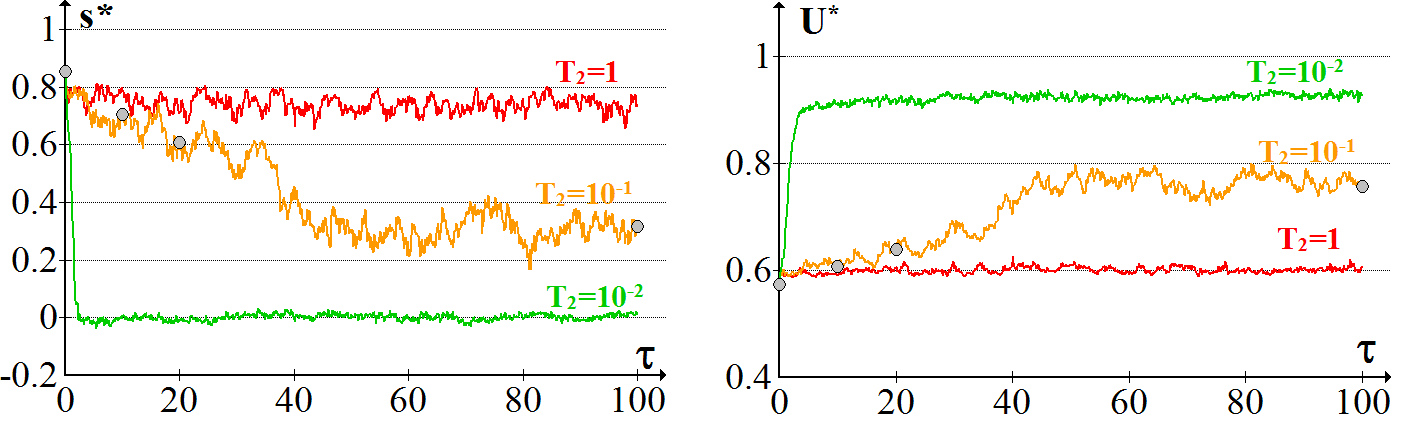}
\end{center}
\caption{\footnotesize{The introduction of local coordination destabilizes the highly segregated configurations and leads to configurations which present locally mixed patterns. \textbf{Top panel}: some snapshots of the evolution of the city for $T_2=0.1$ and $h=4$. \textbf{Bottom panel}: evolution with $\tau$ of the index of similarity and of the collective utility, the grey dots on the $T_2=0.1$ curves corresponding to the snapshots of the top panel. $T=0.1$ and $m=0.5$.}}
\label{QV}
\end{figure}

\indent We present on fig \ref{QV} a typical evolution of the city in the case where the agents move according to the local coordination rule by consulting before each move $h=4$ out of $H=8$ of their potentially new co-proprietors. The level of noise (attached to the co-proprietors) is fixed to $T_2=0.1$ (in order to be comparable to the chosen value of $T$) and the parameter $m$ of the utility function is fixed to $0.5$. Starting from a highly segregated configuration, we observe the disaggregation of the two large homogeneous areas into a much less segregated configuration presenting more local patterns of segregation. This first simulation hence shows that the introduction of a bit of local coordination can be sufficient to break undesired segregated patterns and therefore, \emph{a fortiori}, to prevent segregation to appear starting from a mixed configuration.\\ 

\indent The explanation of why local coordination works is quite simple. 
It corrects the default of the WC mechanism by rendering highly segregated configurations less stable than before since if a red agent goes by mistake into the green area, he will, compared to the WC case, encourage a second red agent to join him. 
Hence the formation of nuclei is encouraged by local coordination. 
A second kind of mechanism to get out of segregated patterns is the advance of the frontier zone.
On the opposite, the locally mixed patterns are more stable. 
Indeed, once an integrated pattern is reached, the co-proprietors tend to prevent the moves which would increase local segregation. \\

\indent It remains to test the robustness and the limit of this results by varying the different parameters.\\

\subsubsection{Influence of the parameters $T_2$ and $m$}
\begin{figure}[h!]
\begin{center}
\includegraphics[width=6.5cm]{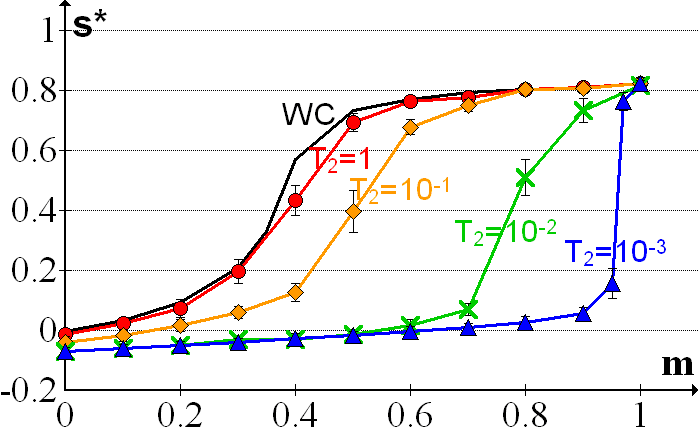}
\includegraphics[width=6.5cm]{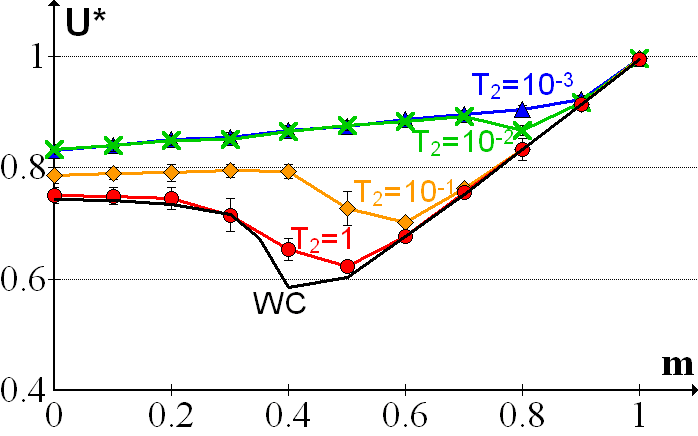}
\end{center}
\caption{\footnotesize{\textbf{Stationary mean values of the similarity index and of the collective utility} as a function of $m$, for different level of noise $T_2$, with the qualified vote dynamic rule. The error bars give the standard deviation of the fluctuations of $s^*$ and $U^*$ once the system has reached its stationary phase. The mean value and standard deviation are computed over $10$ periods of the stationary phases. The plots corresponds to  a neighborhood size $H=8$, a co-properties size $h=4$ and a noise level $T=0.1$. The black curve corresponds to 'Without Coordination' simulations that have been performed using the same parameters.}}
\label{QV2}
\end{figure}

\indent The influence of the parameters $T_2$ and $m$ can be observed on Figs. \ref{QV} to \ref{QVstudy}. For high values of $T_2$, the co-proprietors decision whether to accept or not the moving agent is purely random and the local coordination mechanism has no impact on the dynamics. Hence on fig \ref{QV2} the values of $s^*$ and $U^*$ are similar to their values in the 'without coordination' case for $T_2\gg 1$. On the contrary, when $T_2$ is lowered, the local coordination mechanism allows to break the segregation patterns more rapidly (according to the bottom panel of fig \ref{QV}), leading to mixed configurations presenting locally ordered patterns (according to left panel of fig \ref{QVstudy}). Notice on Fig \ref{QV2} that the impact of local coordination - in terms of welfare issues - is more important for middle values of $m$ (between $0.2$ and $0.8$), precisely the values for which the WC is the most deficient when compared to random allocations.   

\subsubsection{Influence of $h/H$}
\begin{figure}[h!]
\begin{center}
\includegraphics[width=6.5cm]{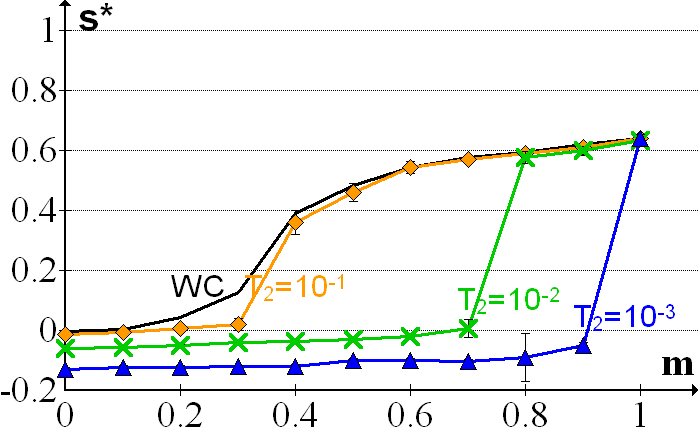}
\includegraphics[width=6.5cm]{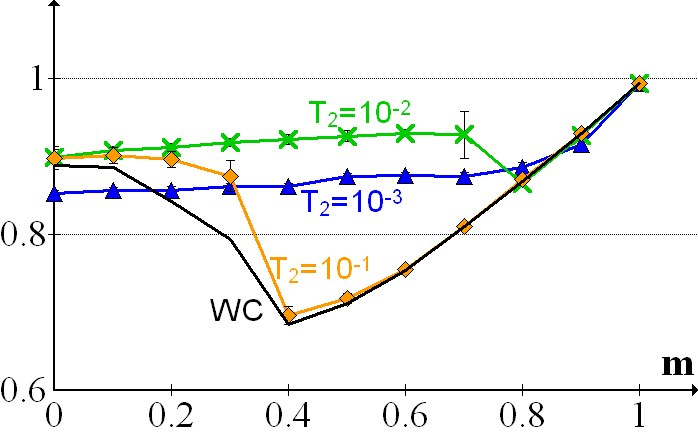}
\end{center}
\caption{\footnotesize{\textbf{Stationary mean values of the similarity index and of the collective utility} as a function of $m$, for different level of noise $T_2$, with the qualified vote dynamic rule. The error bars give the standard deviation of the fluctuations of $s^*$ and $U^*$ once the system has reached its stationary phase. The mean value and standard deviation are computed over $10$ periods of the stationary phases. The plots correspond to a neighborhood size $H=44$, a co-properties size $h=4$ and a noise level $T=0.1$. The black curve corresponds to 'Without Coordination' simulations that have been performed using the same parameters.}}
\label{QV3}
\end{figure}

\begin{figure}[h!]
\begin{center}
\includegraphics[width=6.5cm]{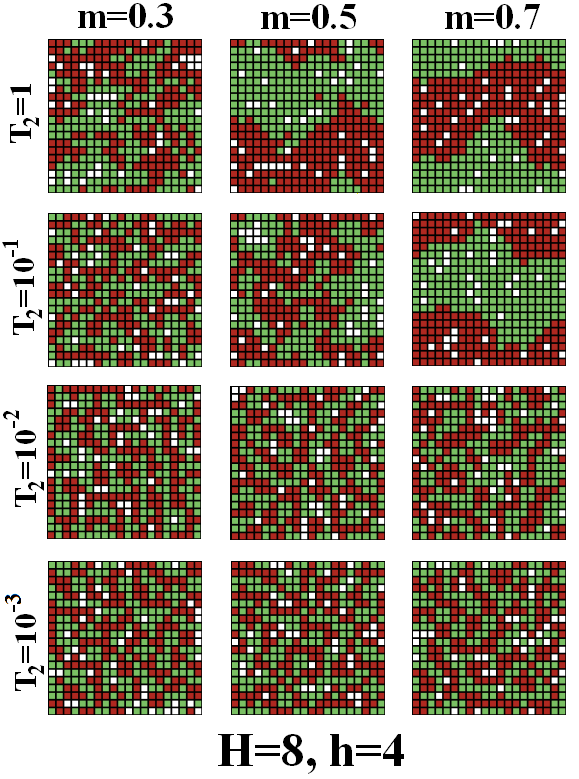}
\includegraphics[width=6.5cm]{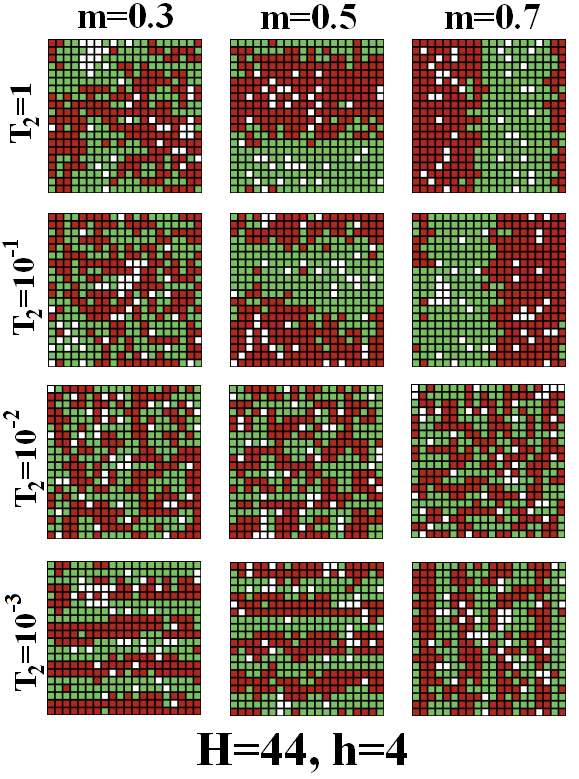}
\end{center}
\caption{\footnotesize{\textbf{Snapshots of typical stationary configurations obtained with the  ``qualified vote'' dynamic rule} for different values of $m$ and $T_2$. \textbf{Left panel}: the $h=4$ new co-proprietors of a moving agent out of the $2H=16$ agents he potentially affects by moving are consulted. \textbf{Right panel}: the $h=4$ new co-proprietors of a moving agent out of the $2H=88$ agents he affects by moving are consulted. Noise level: $T=0.1$.}}
\label{QVstudy}
\end{figure}

The results displayed on Fig \ref{QV3} and on the right panel of Fig \ref{QVstudy} correspond to simulations where the moving agents consult $h=4$  new co-proprietors out of $H=44$ of their potentially neighbors. Since the move of an agent can affect at most $2H$ agents (neighbors in the departure and arrival locations), the LC mechanism is only taking care of $h/2H \simeq 5\%$ of the agents affected by the externalities generated by the moving agents\footnote{In fact, since some of the potentially new neighbors share almost the same neighborhood than voting co-proprietors, there are spatial correlations between the $H$ potentially new neighbors. Hence the LC mechanism takes effectively into account more than $h/2H$ of the affected agents.}. 

For $T_2\geq 10^{-2}$, the results are comparable to the previous one: even if the size of the co-property is relatively less important, the LC mechanism still allows to break the segregated patterns and lead to locally mixed and ordered patterns. The size $H$ of an agent's neighborhood being greater than previously, the typical size of these ordered patterns is also greater.\\

\indent For $T_2=10^{-3}$ however, one can observe that $U^*$ is lower than in the $T_2=10^{-2}$ case (Fig \ref{QVstudy}), and that the normalized similarity $s^*$ is negative for $m\leq0.9$, meaning that the agents have on the average less similar neighbors than dissimilar ones. The corresponding snapshots of the stationary configurations on Fig \ref{QV3} show that the system ends in stripe-like globally ordered states. This result can be understood through the notion of externalities and lack of coordination. Indeed, when $h/H\ll 1$, we can separate the voting co-proprietors whose neighborhood is close to the vacant cell envisaged by the moving agent and the neighbors living on a further ring. The respective neighborhoods of these two kind of neighbors are not spatially correlated, which means that the interest of these two kind of neighbors are clearly different. There are hence three kind of agents at play: the moving one, the inner ring of new neighbors (among which the voting agents belong) and all the other affected agents (the outer ring of new neighbors and the former neighbors). For finite values of $T$ and $T_2\rightarrow 0$, the interest of the second group is in fact the sole taken into account in the LC mechanism. There is no effective coordination between all the involved agents, some externalities created by the moves are not taken into account and the system gets away from welfare maximizing states.     

\section{Discussion}
The aim of the work presented in this chapter is to study the effects of the introduction of coordination in a Schelling model. 
We showed here that introducing partial coordination through a tax on the externalities generated by the individual move is sufficient to break the gap between the agents' micro-motives and the emergent macro behavior and therefore to break undesired segregative patterns. Moreover, we showed that it is not necessary to tax all the externalities to reduce it significantly. A tax equivalent to one fifth of the externalities might well be sufficient.\\ 

In a second mechanism, we introduced local coordination through a voting mechanism which involves only individual decisions and does not require the intervention of a benevolent central authority as in the tax mechanism. We show that in cases when the vote mechanism allows to take into account the interest of a large enough part of the agents affected by a move, this mechanism is sufficient to reduce significantly the above mentioned gap and reach a high level of collective utility.\\

In regard to segregation issues, the introduction of coordination cannot break the segregative pattern without the agents having a certain preference for mixed neighborhood. Coordination is only a way to reinforce on the large scale the wishes of the agents on the local scale: if they are intolerant, segregation will occur, if they are tolerant, integration may occur. Another issue that is only partly addressed here are the typical times needed to break the segregated configuration and particularly the influence of the initial configurations in this matter. We leave that point to further studies.\\

\chapter*{Conclusion}
\addcontentsline{toc}{chapter}{Conclusion}

\section*{Overview of the paper}
We gave in Chapter $1$ an overview of the concepts at stake in Schelling models, namely neighborhood description, individual preferences and dynamic rules. We particularly pointed out the relative efficiency (in terms of statistical significance and reproducibility of the outcomes) of the logit dynamic rule compared to a Best-Response dynamic rule.\\

In chapter $2$, we used recent tools from evolutionary game theory to develop an analytical resolution valid in a particular context, i.e. bounded neighborhoods and two homogeneous groups of agents. This is a significant step forward compared to previous analytical work, mostly limited to computer simulations. We showed that the stationary configurations ensuing from the selfish individual moves of the agents maximize a potential function that can be interpreted in terms of the sum of the agents' utilities at the time when they have moved into their current location starting with a totally empty city,
that is in other words, the incentives the agents had to move in the neighborhood where they are located.
Thanks to this potential function, we analyzed several versions of Schelling models with different utility functions. In particular, when the agents' utility function is Schelling's original utility function (stair-like function with a threshold at $s=0.5$), the potential function becomes the Duncan and Duncan segregation index.\\

Finally, in chapter $3$, we presented extended versions of Schelling-type models incorporating different kinds of coordination between the agents. More particularly, local coordination through a vote mechanism, while remaining in the ``individual decision'' spirit of Schelling model, is shown to be sufficient to reach stationary configurations with a significantly higher collective utility than a standard Schelling model. We hope that such an individualistic based coordination model can be seen as a valuable alternative to the coordination models presented in the literature, mostly based on affirmative action policies \citep{dokumaci07}.  

\section*{Explaining the emergence of segregation}
\indent We have verified, both by simulations and by analytical calculations, the paradoxical result that has generated interest for Schelling's model. While the dynamics is governed by agents moving to improve their own utility, their moves lead to highly segregated configurations at the city level in which most of the agents are far from being fully satisfied. Our paper shows that asymmetry in the utility function - that is a slight taste for like-neighbors - and selfish behavior are the two main ingredients of segregation\\

\indent The externalities generated by selfish moves are the important ingredient hightlighted in previous literature. 
Actually, as already argued by \cite{zhang2004rsa} and \cite{pancs07}, individual preferences for integrated environments may lead to segregated configurations because location choice by an agent affects her neighbors' utility. 
We stress why this makes mixed neighborhoods unstable and segregated configurations very stable. 
The instability of mixed neighborhoods is particularily clear in the block configuration for the asymmetrically peaked function. 
Starting with the Nash equilibrium of a perfectly mixed neighborhood, there is a positive probability that an agent accepts a slight decrease of its utility, and leaves this block. 
The agents of the same colour remaining in the block now have a lower utility and are even more likely to leave. 
This creates an avalanche which empties the block, as each move away further decreases the utility of the remaining agents. 
Conversely, highly-segregated configurations are very stable. Indeed, once the city is divided into homogeneous areas, a red agent will have no incentive to go from the red area to the green one (his utility dropping from $0.5$ to $0$). \\

As stated by \cite{zhang2004rsa}: ``\emph{although nobody likes complete segregation, the residential pattern is very stable. 
Only moving across the color line by a considerable number of agents could disturb the segregation equilibrium, but nobody has incentive to do so because it causes a loss of [individual] utility.
 [...] Segregation is stable not because people like it, but because any individual who wants to change the situation unilaterally will have to go across the color line, which may not be the desirable thing to do from the individual's perspective.
 The failure of the system to escape complete segregation is similar to the phenomenon of  ``coordination failure'' studied by economists in many other contexts. It is the agents' inability to move simultaneously that make them stuck in a situation nobody likes...}''\\

\indent The most important element driving segregation is the {\it asymmetry} of the utility function. 
Symmetric functions do not lead to segregation. It is only if utility functions favor a large-majority status over a small-minority status,
that segregation is found in spite of a strict preference for mixity, as in the asymmetrically peaked function.
Beyond this particular example, our analytical analysis gives a general rule in terms of sufficient conditions on the utility functions for steady-state configurations to be segregated. 
This rule opens the path for analyzing more carefully different utility functions and their relations to households' preferences.  \\

\section*{Future work}
 
One of the most interesting tracks for future work would certainly be to explore more thoroughly what drives the agents' preferences regarding their neighbors' attributes. 
Indeed, in real life preferences regarding mixity seem influenced by individual past experiences as well as social norms.
Such an analysis could be done by coupling Schelling's model with another model describing the dynamic evolution of preferences.
For example, one could introduce heterogeneity in the agents' preferences and allow them to evolve over time, taking into account the individual's past experiences and its neighbors' preferences.


\appendix
\chapter{Appendix}

\section{Additions to chapter $1$}

\subsection{Justification of the logit rule}
\label{Alogit}
\indent The logit rule can be explicitly derived within a random utility model. In this context, the utility $\tilde{u}$ of an agent is composed of two terms: a deterministic term $u$ which depends on his neighborhood's composition and a random term $\epsilon$ which is introduced to take into account any other characteristics of the location and its neighborhood which the agent may be sensitive to (the proximity from the city center, school or supermarket, etc...). The random term $\epsilon$ is assumed to be independent both across agents and locations. Agent $k$'s total payoff on location $i$ is thus written as
\begin{equation*}
\tilde{u}_{k,i}=u_{k,i}+T\epsilon_{k,i}
\end{equation*}  
\indent The parameter $T$ is a positive constant which determines the relative importance of the random term. If $T$ is close to $0$, the random term is not important and can be neglected. If it is close to infinity, the random term is very important and the neighborhood's compositions do not play any role. Here, neighborhood's composition is supposed to be the main determinant of agents' actions : we restrict our analysis to the case of low values of $T$.\\
\indent The actions of each agent being solely based on their own profit, the potential mover will choose to move from location $i$ to location $j$ if and only if it increases his utility $\tilde{u}$, ie iff $\,u_{k,i}+T\epsilon_{k,i} < u_{k,j}+T\epsilon_{k,j}$. Following \citep{mcfadden73}, we assume that the $\epsilon_{k,i}$ are independent and follow identical extreme value distribution whose cumulative distribution function and probability distribution function are
$$F(x) = \exp(-e^{-x}),\,\, f(x) = exp(-x - e^{-x})$$
\indent Then, noting $\Delta u=u_{k,j}-u_{k,i}$ the gain the potential mover would achieve if he was to move and $\phi=e^{-\epsilon}$, 
\begin{eqnarray}
Pr\{move\} &=& Pr\{u_{k,i} +T\epsilon_{k,i} < u_{k,j}+T\epsilon_{k,j}\} \,=\, Pr\{\epsilon_{k,i} < \Delta u/T + \epsilon_{k,j}\}\nonumber\\
&=& \int_{-\infty}^{+\infty}\,F(\Delta u/T + \epsilon )\cdot f(\epsilon)d\epsilon \,=\, \int_{-\infty}^{+\infty}\,\exp( -e^{-\Delta u/T - \epsilon})\cdot exp(-\epsilon - e^{-\epsilon})d\epsilon \nonumber \\ 
 &=& \int_{-\infty}^{+\infty}\,exp\Big(-\epsilon - e^{-\epsilon}(1+e^{-\Delta u/T_1})\Big)d\epsilon \,=\, \int_{0}^{+\infty}\,e^{-\phi(1+e^{-\Delta u/T})}d\phi \nonumber \\
 &=& (1+e^{-\Delta u/T})^{-1}\int_{0}^{+\infty}e^{-\phi}d\phi \,=\, (1+e^{-\Delta u/T})^{-1}\cdot 1 \nonumber 
\end{eqnarray}
\indent The probability that the potential mover chooses to move is thus determined by:
$$Pr\{move\} = \frac{1}{1+e^{\,-\Delta u / T}} \hspace{0.7cm}\square $$
\indent The random term $\epsilon$ in the payoff function can alternatively be interpreted as a way to model the agents' bounded rationality: it may happen that an agent takes a utility-decreasing move, either because he is making a mistake or because of a lack of information.\\

\subsection{Proof of Lemma 1}
\label{Alemma1}
\indent For any two states $x$ and $y$, let $P_{yx}$ be the transition probability from state $y$ to state $x$ and $t_{yx}$ the expected number of iterations for the system, starting from state $y$, to arrive at state $x$. Starting from state $y$, let us assume $1$ iteration has passed and examine the expected number of additional time periods it takes for the system to arrive at state $x$. With probability $P_{yx}$ the system is already in $x$ and it takes $0$ additional time periods. With probability $P_{yk}$ the system is in state $k\neq x$, and it takes $t_{kx}$ periods to arrive at $x$. Therefore, $t_{yx}$ can be written as 
$t_{yx}=1+P_{yx}\cdot 0 + \sum{k\neq x}P_{yk}t_{kx}=1+\sum{k\neq x}P_{yk}t_{kx}$
Multiply both sides of the equation by the stationary probability $\Pi(y)$ and sum over all states to obtain 
\begin{eqnarray*}
\sum_{y\in X}\Pi(y)t_{yx} &=& \sum_{y\in X}\Pi(y) + \sum_{y\in X}\sum{k\neq x}\Pi(y)P_{yk}t_{kx}\\
&=& 1 + \sum{k\neq x}t_{kx}\sum_{y\in X}\Pi(y)P_{yk} = 1+ \sum{k\neq x}t_{kx}\Pi(x)
\end{eqnarray*}

This implies that 
$1=\sum_{y\in X}\Pi(y)t_{yx}-\sum_{k\neq x}m_{kx}\Pi(x)= \Pi(x)t_{xx}$. It immediately follows that
$\Pi(x)=1/t_{xx}$.

\subsection{Robustness with the neighborhood size}
\label{Aneigh}
\begin{figure}[h!]
\begin{center}
\includegraphics[width=13cm]{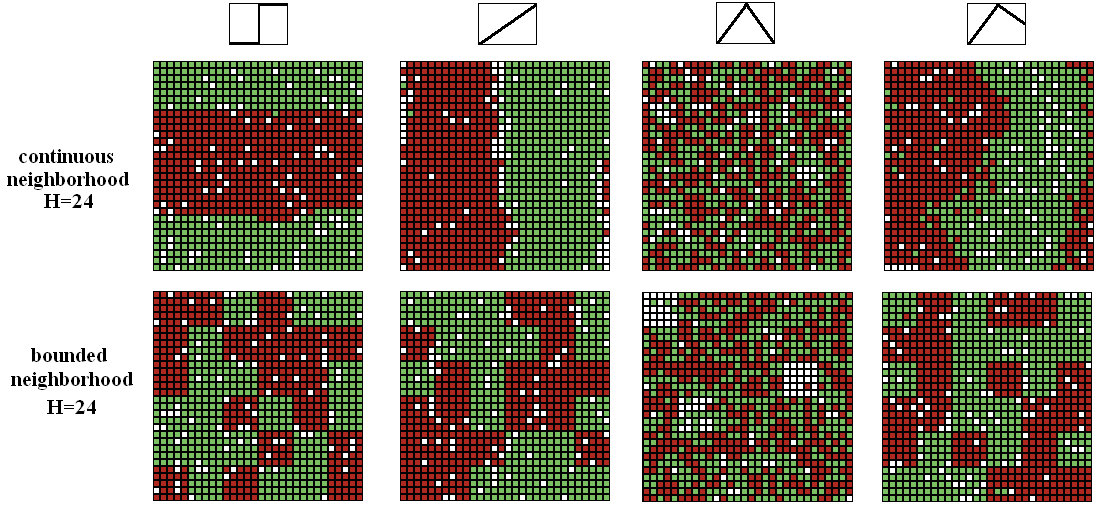}
\end{center}
\caption{\footnotesize{Typical stationary configurations obtained by simulations for different models. The demographic parameters are $(N=30,v=10\%,n_R=0.5)$. Neighborhood sizes are fixed to $H=24$. Up: with a continuous neighborhood description. Down: with a bounded neighborhood description. From left to right: the agents compute their utility with some of the functions presented Fig \ref{util}. The level of noise is fixed to $T=0.1$}}
\label{influNu2}
\end{figure}

\indent Results presented on fig \ref{influNu2} are similar to the ones presented on Fig \ref{influNu}, the difference being that the neighborhood size is fixed here at $H=24$ instead of $H=8$.\\

\indent We verify once again that the qualitative behavior of the system does not depend on the choice of neighborhood description. Another important fact is that the qualitative outcomes presented here are the same than those for $H=8$. We do not find any qualitative difference for other values of $H$ (not shown here), provided $H\ll N^2$.\\


\section{Additions to chapter $2$}

\subsection{Proof of Claim 1}
\label{proof1}
indent Let us first prove the first part of claim $1$, that is that any aggregate function $\mathcal{F} = \sum_{q\in\mathcal{Q}}F(R_q,G_q)$ is a potential function that corresponds to (at least) one pair of utility functions $(u_R,u_G)$ of $\mathbb{U}$.\\ 
\indent Suppose that $\mathcal{F}=\sum_{q\in\mathcal{Q}}F(R_q,G_q)$ is a potential function of the game, where the intermediate function $F$ is known. Let us assume that an agent is moving from a block $1$, characterized by the numbers $(R_1,G_1)\in E_{H+1}$ of red and green agents that live in it, to a block $2$ characterized similarly by the numbers $(R_2,G_2)\in E_{H}$ of red and green agents that live in it (since there must be at least one vacant location in block $2$ for an agent to move in it, we necessarily have $R_2+G_2<H+1$). By definition, the utility variation of a moving agent must be equal to the variation of $\mathcal{F}$ it induces. Hence :\\
\indent $\bullet$  to cover the cases when the moving agent is a red one it is necessary that for all $(R_1,G_1)\in E_{H+1}$ with $R_1 \geq 1$,
\begin{eqnarray}
u_R(R_2,G_2) &-& u_R(R_1-1,G_1) = \nonumber\\
&& F(R_2+1,G_2) + F(R_1-1,G_1) - F(R_2,G_2) - F(R_1,G_1) \label{eqB1}
\end{eqnarray}
\indent $\bullet$ to cover the cases when the moving agent is a green one it is necessary that for all $(R_1,G_1)\in E_{H+1}$ with $G_1 \geq 1$,
\begin{eqnarray}
u_G(R_2,G_2) &-& u_G(R_1,G_1-1) = \nonumber\\
&& F(R_2,G_2+1) + F(R_1,G_1-1) - F(R_2,G_2) - F(R_1,G_1) \label{eqB2}\\\nonumber
\end{eqnarray}

\indent Considering the case where $R_2=G_2=0$, equations \ref{eqB1} and \ref{eqB2} can be rewritten so that one finds that the utility functions $u_R$ and $u_G$ verify for all $(R,G) \in E_{H}$:
\begin{eqnarray}
u_R(R,G) - u_R(0,0) &=& F(R+1,G)-F(R,G) - F(1,0) + F(0,0)\label{eqB3}\\
u_G(R,G) - u_G(0,0) &=& F(R,G+1)-F(R,G) - F(0,1) + F(0,0)\label{eqB4}
\end{eqnarray}
\indent These relations define (up to a constant $u(0,0)$) the utility functions the agents necessarily have if  $\mathcal{F}=\sum_{q\in\mathcal{Q}}F(R_q,G_q)$ is a potential function of the game. It still remains to prove that this pair of utility functions belongs to the set $\mathbb{U}$. According to the relations \ref{eqB3} and \ref{eqB4}, one have for all $(R,G)\in E_H$:
\begin{eqnarray*}
 u_R(R,G) - u_R(R,G+1) &=& \big(F(R+1,G) - F(R,G)\big) - \big(F(R+1,G+1) - F(R,G+1)\big)\\
 &=& \big( F(R,G+1) - F(R,G)\big) - \big( F(R+1,G+1) - F(R+1,G)\big) \\
 &=& u_G(R,G) - u_G(R+1,G)
\end{eqnarray*} 
\indent Hence the relation \ref{cond} holds, which means by definition that the pairs of utility functions $(u_R,u_G)$ defined by the relations \ref{eqB3} and \ref{eqB4} belongs to $\mathbb{U}$.\\\\

\indent Let us now prove the second part of claim $1$, which is that to any pair of utility functions $(u_R,u_G)$ of $\mathbb{U}$ corresponds a potential function of the form $\mathcal{F} = \sum_{q\in\mathcal{Q}}F(R_q,G_q)$.\\
\indent Let $(u_R,u_G) \in \mathbb{U}$ be a pair of utility functions. Suppose that $F(0,0)$, $F(0,1)$ and $F(1,0)$ are given and let us define recursively the function $F$ on $E_{H+1}$ by the following equations, verified for all $(R,G) \in E_{H}$:
\begin{eqnarray}
F(R+1,G) - F(R,G)  &=& F(1,0)  - F(0,0) + u_R(R,G) - u_R(0,0) \label{eqB5}\\
F(R,G+1) - F(R,G)  &=& F(0,1) + F(0,0) + u_G(R,G) - u_G(0,0) \label{eqB6}
\end{eqnarray}

\indent The most important thing to notice is that these two relations are consistent with each other thanks to the condition \ref{cond} that links the two utility functions $u_R$ and $u_G$. By summing Eq. \ref{eqB5} on $R$ then Eq.\ref{eqB6} on $G$, one then finds the following expression of the function $F$:
\begin{eqnarray*} F(R,G)-F(0,0) = R\Big( F(1,0)-F(0,0)\Big) + \sum_{r=0}^{R-1} \Big( u_R(r,0) - u_R(0,0)\Big)\\ 
+ G\Big( F(0,1)-F(0,0) \Big)+ \sum_{g=0}^{G-1} \Big( u_G(R,g) - u_G(0,0)\Big)  \\ 
\end{eqnarray*}
or conversely by summing Eq. \ref{eqB6} on $G$ then Eq.\ref{eqB5} on $R$,
\begin{eqnarray*} F(R,G)-F(0,0) = R\Big( F(1,0)-F(0,0)\Big) + \sum_{r=0}^{R-1} \Big( u_R(r,G) - u_R(0,0)\Big) \\
+ G\Big( F(0,1)-F(0,0)\Big) + \sum_{g=0}^{G-1} \Big( u_G(0,g) - u_G(0,0)\Big) \\
\end{eqnarray*}

\indent Hence, since $\mathcal{F} = \sum_{q\in\mathcal{Q}}F(R_q,G_q)$ one obtains, after rearranging the different terms, a symmetric expression of the potential:  
\begin{eqnarray}
\mathcal{F} &=& |\mathcal{Q}|F(0,0) + N_R\big(F(1,0)-F(0,0)-u_R(0,0)\big) + N_G\big(F(0,1)-F(0,0)-u_G(0,0)\big)\nonumber \\
&& \hspace{10mm}+ \frac{1}{2} \sum_{q\in\mathcal{Q}} \Big[\sum_{r=0}^{R_q-1} \Big(u_R(r,0)+u_R(r,G_q)\Big) + \sum_{g=0}^{G_q-1} \Big(u_G(0,g)+u_G(R_q,g)\Big)\Big]
\label{gen-form-F}
\end{eqnarray}

\indent Since the potential can be chosen up to a constant, it is clear from the previous expression that the choice of $F(0,0)$, $F(0,1)$, $F(1,0)$, $u_R(0,0)$ and $u_G(0,0)$ do not really matter. Hence our choice to put them to zero to simplify the generic expressions of the potential given in Eq. \ref{Fu1} and \ref{Fu2}.

\subsection{Calculation of a particular potential function}
\label{appli1}
Suppose that the agents compute their utility with Schelling's utility function (which is equal to $1$ if their fraction of similar neighbors is superior or equal to $0.5$, and equal to $0$ otherwise). This utility function can be expressed in terms of the number of red and green neighbors as follows:
\begin{eqnarray} 
u_R(R,G) &=& \Theta(R-G) = \frac{1}{2}(1+|R+1-G|-|R-G|)\nonumber\\   
u_G(R,G) &=& \Theta(G-R) = \frac{1}{2}(1+|R-1-G|-|R-G|)\label{usch}
\end{eqnarray}
where $\Theta$ is the Heaviside function defined by: $\Theta(x)=0 \,\,\mbox{if } x < 0$ and $\Theta(x)=1 \,\,\mbox{if } x \geq 0$. Notice that in this example (and in this example only) the convention $u(0,0)$ used in claim 1 is not respected. The form we choose to write Schelling's utility function imposes $u_R(0,0)=u_G(0,0)=1$. It is easy to figure out that this particular pair of utility functions respect the condition \ref{cond}, and is therefore in the set $\mathbb{U}$. Indeed, 

\begin{eqnarray*}
u_R(R,G) - u_R(R,G+1) = \Theta(R-G)-\Theta(R-G-1) =
\left\{ \begin{array}{lll} 
  0-0 &= 0 & \mbox{if $ R \leq G-1$ } \\ 
  1-0 &= 1 & \mbox{if $R = G $ }  \\
  1-1 &= 0 & \mbox{if $ R \geq G+1$ } \\ 
\end{array} \right.
\end{eqnarray*}
and
\begin{eqnarray*}
u_G(R,G) - u_G(R+1,G) = \Theta(G-R)-\Theta(G-R-1) =
\left\{ \begin{array}{lll} 
  1-1 &= 0 & \mbox{if $ R \leq G-1$ } \\ 
  1-0 &= 1 & \mbox{if $R = G $ }  \\
  0-0 &= 0 & \mbox{if $ R \geq G+1$ } \\ 
\end{array} \right.
\end{eqnarray*}
Hence the relation $u_R(R,G) - u_R(R,G+1) = u_G(R,G) - u_G(R+1,G)$ is always verified.\\

\indent To compute a corresponding potential function, one can refer to the general form of Eq. \ref{gen-form-F} (since we do not use the convention $u(0,0)=0$ in this partiucular example) which can be written here as:

\begin{eqnarray*}
\mathcal{F} &=& const + \frac{1}{2} \sum_{q\in\mathcal{Q}} \Big[\sum_{r=0}^{R_q-1} \Big(u_R(r,0)+u_R(r,G_q)\Big) + \sum_{g=0}^{G_q-1} \Big(u_G(0,g)+u_G(R_q,g)\Big)\Big] \\
&=& const + \frac{1}{4} \sum_{q\in\mathcal{Q}} \Big[\sum_{r=0}^{R_q-1} \Big(3+|r+1-G_q|-|r-G_q|\Big) + \sum_{g=0}^{G_q-1} 
\Big(3+|R_q-1-g|-|R_q-g|\Big)\Big] \\
&=& const +  \frac{1}{4} \sum_{q\in\mathcal{Q}} \Big[\Big( 3R_q + |R_q-G_q| - G_q\Big) + \Big( 3G_q + |R_q-G_q|-R_q\Big)\Big] \\
&=& const +  \frac{1}{2}\sum_{q\in\mathcal{Q}}\Big( R_q + G_q + |R_q-G_q|\Big) \\
&=& const + \frac{1}{2}(N_R + N_G) + \frac{1}{2}\sum_{q\in\mathcal{Q}}|R_q-G_q| \\
&=& const' +  \frac{1}{2}\sum_{q\in\mathcal{Q}} |R_q-G_q|
\end{eqnarray*}

It is also possible to compute the potential function more directly thanks to the interpretation we presented in section \ref{interpr}. The potential function corresponds to the sum of the utility of the agents being introduced one by one in the city, this sum being independent of the precise order of the introduction of the agents. To compute the potential of a given configuration $\{R_q,G_q\}$, let's consider that we introduce in each block first the agents in majority (ie the red ones if $R_q > G_q$, the green ones if $G_q> R_q$, either the red or the green ones if $R_q=G_q$) and second the agents in minority. Each of the first agents has a utility of $1$ as he settles in the city (since his group is in majority in his block when he settles) while each of the other agents has an zero utility when he settles (since his group is in minority when he settles). Hence it is straightforward to write the potential as

\begin{eqnarray*}
\mathcal{F} &=& const + \sum_q \max(R_q,G_q)\\
&=& const +  \sum_{q\in\mathcal{Q}}\frac{1}{2}\Big( R_q + G_q + |R_q-G_q|\Big) \\
&=& const' +  \frac{1}{2}\sum_{q\in\mathcal{Q}} |R_q-G_q|
\end{eqnarray*}

This first example illustrates the usefulness of the interpretation of the potential function as the sum of the agents (introduced one by one) settling's utilities. The computation of $\mathcal{F}$ is indeed much easier and bears more meanings with the second method. To compute the potential function corresponding to a given pair $(u_R,u_G)$ of utility functions, it may be worth to think ahead of a practical order of introduction
of the agents.

\subsection{Relation between the potential function $\mathcal{F}$ and the collective utility $U$ }
\label{proof_FU}
\indent Let us suppose that $(u_R,u_G)\in \mathbb{U}$, and that the potential function of the system can be expressed as a linear function of the collective utility, \emph{ie} $\mathcal{F}(\{R_q,G_q\})= \lambda U(\{R_q,G_q\}) + \mu$. Since the potential function can be defined up to constant, we can take $\mu=0$. 
Writing the utility functions under the form
\begin{eqnarray*}   
u_R(R,G) &=& \xi_R(R) + \sum_{g=0}^{G-1} \xi(R,g) \\
u_G(R,G) &=& \xi_G(G) + \sum_{r=0}^{R-1} \xi(r,G) 
\end{eqnarray*}
introduced in Eq. \ref{cond2a} and \ref{cond2b}, the relation of proportionality between the potential and the collective utility can be written as 
\begin{eqnarray*} 
&&\sum_q\Big(\sum_{r=0}^{R_q-1} \xi_R(r) + \sum_{g=0}^{G_q-1} \xi_G(g) + \sum_{r=0}^{R_q-1} \sum_{g=0}^{G_q-1} \xi(r,g)\Big) \\
&=& \lambda \sum_q\Big(R_q \xi_R(R_q-1) + R_q \sum_{g=0}^{G_q-1}\xi(R_q-1,g) +G_q \xi_G(G_q-1) + G_q \sum_{r=0}^{R_q-1} \xi(r,G_q-1)\Big)
\end{eqnarray*} 

Since this relation must hold for all $\{R_q,G_q\}$, it follows that that for all $(R,G) \in E_H$, the following holds: 

\begin{eqnarray} 
&&\sum_{r=0}^{R-1} \xi_R(r) + \sum_{g=0}^{G-1} \xi_G(g) + \sum_{r=0}^{R-1} \sum_{g=0}^{G-1} \xi(r,g) \nonumber\\
&=& \lambda \Big(R \xi_R(R-1) + R \sum_{g=0}^{G-1}\xi(R-1,g) +G \xi_G(G-1) + G \sum_{r=0}^{R-1} \xi(r,G-1)\Big)
\label{rel}
\end{eqnarray}

Taking successively $G=0$ and $R=0$ in that last equation provides three independent relations dissociating the three functions $\xi_R$, $\xi_G$ and 
$\xi$:

\begin{eqnarray}
\forall R>0, &\sum_{r=0}^{R-1}\xi_R(r) = \lambda R\xi_R(R-1) \label{AA1}\\
\forall G>0, &\sum_{g=0}^{G-1}\xi_G(g) = \lambda G\xi_G(G-1) \\
\forall (R,G), \in E_H &\sum_{r=0}^{R-1}\sum_{g=0}^{G-1}\Big(\lambda \xi(R-1,g) + \lambda\xi(r,G-1) - \xi(r,g)\Big) = 0 \label{AA2}
\end{eqnarray}   

Notice moreover that the convention $u(0,0) = 0$ implies $\xi_R(0) = \xi_G(0) = 0$. Let us also suppose $\xi_R(1)=a \neq 0$, $\xi_G(1)=d\neq 0$ and $\xi(0,0)=b$. Starting from equations \ref{AA1} to \ref{AA2}, it is straightforward to prove recursively that    
\begin{eqnarray*}
&\lambda = 1/2 \\
\forall R>0, &\xi_R(R) = aR \\
\forall G>0, &\xi_G(G) = dG \\
\forall (R,G) \in E_H &\xi(R,G) = b
\end{eqnarray*}

Hence the agents' utility functions corresponds exactly to those introduced in Eq. \ref{uabd}:
\begin{eqnarray*}   
u_R(R,G) &=& aR +bG \\
u_G(R,G) &=& bR + dG 
\end{eqnarray*}

The individual utilities are thus necessarily linear in the numbers of similar and dissimilar neighbors in case the potential function $\mathcal{F}$ is proportional to the collective utility $U$ $\square$



\end{document}